\documentclass[12pt,preprint]{aastex}

\shorttitle{Line survey of NGC\,7027}
\shortauthors{Zhang \& Kwok}

\begin{document}

\title{A spectral line survey of NGC\,7027 at millimeter wavelengths}

\author{Yong Zhang \& Sun Kwok} 
\affil{Department of Physics, University of Hong Kong, Pokfulam Road, Hong Kong}
\email{zhangy96@hku.hk; sunkwok@hku.hk}

\author{Dinh-V-Trung}
\affil{Institute of Astronomy and Astrophysics, Academia Sinica\\ 
P.O Box 23-141, Taipei 106, Taiwan}
\email{trung@asiaa.sinica.edu.tw}

\begin{abstract} 

We report on a recent 
spectral line survey of the planetary nebula (PN) \object{NGC\,7027}
using the Arizona Radio Observatory (ARO) 12\,m telescope and 
the Heinrich Hertz Submillimeter Telescope (SMT) at millimeter wavelengths.
The spectra covering the frequency ranges 71--111\,GHz, 157--161\,GHz, and
218--267\,GHz were obtained with a typical sensitivity of $rms<8$\,mK.
{ A total of 67 spectral lines are detected, 21 of which are
identified with 8 molecular species, 32 with recombination lines 
from hydrogen and helium, and 14 remains unidentified. }
%HC$_3$N is clearly detected for the first time in a PN.
%, implying to the presence of dense clumps which shield the UV radiation from the central star.
%Rotation-diagram analysis of HC$_3$N suggests that it arises 
%from two temperature components. 
As the widths of emission lines from CO, other neutral molecules, molecular ions, as well as recombination of H$^+$ and He$^+$ are found to be different with each other, the line strengths and profiles are used to investigate the physical conditions and chemical processes
of the neutral envelope of \object{NGC\,7027}.
The column densities and fractional abundances relative to
H$_2$ of the observed molecular species are calculated and compared with predictions from chemical models. We found evidence for overabundance of N$_2$H$^+$ and underabundance of CS and 
HNC in \object{NGC\,7027}, suggesting that X-ray emission and shock 
wave may play an important role in the chemistry of the hot molecular
envelope of the young PN. 
%The isotopomer ratios of carbon and oxygen, products of nucleosynthesis and dredge-up processes in eolved stars, are found to be consistent with the presence of cold bottom burning.
%We have determined the isotopomer ratios of carbon and oxygen, which enable to study nucleosynthesis and dredge-up processes in evolved stars. Our results support the presence of cold bottom processing (CBP).
%The analysis of radio recombination lines shows that they are optical thin, and stimulated excitation may is important.

\end{abstract}


\section{Introduction}

We now realize that the circumstellar envelopes of evolved stars are active sites of molecular synthesis.  Millimeter-wave spectroscopy has identified over 60 molecular species in the envelopes of asymptotic giant branch (AGB) stars, their descendants planetary nebulae (PNs), and the transition objects between the two phases, proto-planetary nebulae (PPNs) { \citep{olo97,cernicharo00,ziu07}}.  These gas-phase molecules are believed to be the precursors of complex organic compounds that are found in PPNs and PNs { \citep{cernicharo01a,cernicharo01b,kwo04}}.  Since the molecular synthesis takes place in the circumstellar envelope through either gas-phase or grain-surface reactions, the study of the changing chemical composition and molecular abundance between objects in consecutive phases of evolution will provide useful information on the chemical pathways of their formation.
Furthermore, since the circumstellar envelopes are expanding, the dynamical 
timescales impose a time limit on the reaction time scales. { The
typical dynamical timescales for AGB stars, PPNs, and PNs are
$10^4$--$10^5$ yr, $10^3$ yr, and $10^3$--$10^4$ yr, respectively.}
%\citep{mccarthy90,young93,villaver02a,villaver02b}.}
%For AGB stars, 
%the typical dynamical time is $10^4$ yr, and for PNs, $10^4$ yr.  Constraints 
%on the chemical timescale for PPNs can be found in a combination of 
%evolutionary and dynamical timescales, which are of the order of $10^3$ yr.
The observation of a specific molecular species in the PPN phase which is not seen in AGB stars implies that this molecule must be formed on timescales shorter than a few thousand years.  The changing relative abundance of molecular species between different phases can also point to specific chemical pathways.  Motivated by these considerations, we have performed spectral line surveys of several objects in the late stages of stellar evolution and try to provide a systematic and uniform study of circumstellar chemistry.

%Observations of molecules provide an important tool in the study of interstellar chemistry. Spectral line surveys at millimeter wavelengths have been extensively obtained in the past to study the formation of molecules in circumstellar envelopes of late-type stars.   Many new molecular species have been revealed, providing an opportunity to probe physical conditions and gaseous kinematics in the molecular  envelopes. The investigations of the molecular lines from different types of astrophysical sources allow us to understand the chemical processes in different evolutionary stages of stars. Rich molecules can be rapidly formed during the mass lost on the asymptotic giant branch (AGB) stage. Planetary nebulae (PNs) evolve from the envelope of AGB stars and have a hotter central star and lower gaseous density. Unlike the circumstellar envelopes of AGB stars, the chemistry of PNs is  dramatically affected by the strong UV radiation from the central stars. The molecular composition of PN envelopes thus is expected to be different from that in the envelopes of AGB stars. A comparison study of molecules in the envelopes of AGB stars and in those of PNs is vital to understand the chemical evolution in interstellar medium. However, observations of molecular lines in PNs are scarce to date.

The study of the circumstellar chemistry of PNs is particularly interesting and challenging because of the additional dimension of photochemistry.  Since part of the circumstellar envelope is ionized by ultraviolet radiation from { the hot} central star, molecules can be photoionized or photodissociated.  The molecular composition of the circumstellar envelopes is therefore expected to be qualitatively different from AGB stars and PPNs.
The young, carbon-rich PN \object{NGC\,7027} is known to be the PN with the richest molecular content.  Because of its high-mass (0.7 M$_\odot$) central star, the central star of \object{NGC\,7027} is evolving very quickly, reaching an effective temperature of 219\,000\,K in a few hundred years since the onset of photoionization \citep{zhang05}.  %Given  its proximity, high surface brightness and rich emission-line spectrum, \object{NGC\,7027} has been known as an archetypal object for studies of ionized and neutral gas in PNs.
The remnant of the AGB envelope is eroded by a nascent ionized region of 10\arcsec.  Imaging observations have shown that a thin photodissociation region (PDR) separates the ionized region and the extended molecular envelope with a size $>70''$
\citep[see][and the references therein]{hasegawa01}.
A large number of molecular emission lines have been detected by infrared (IR) and radio spectroscopy 
{ \citep[e.g.][]{liu96,cox93,hasegawa01,herpin02}}. Unambiguously
detected molecular species in \object{NGC\,7027} include H$_2$, OH, CO, CO$^+$, CH, CH$^+$,HCN, HNC, HCO$^+$,
H$_2$O, N$_2$H$^+$, CN, CS, C$_2$H, C$_3$H$_2$, SiS, and their 
isotopomers. A systematic molecular-line observations of PNs by \citet{bachiller97} and \citet{josselin03} has found that the column densities of molecules observed in \object{NGC\,7027} are generally larger than those in other PNs.  Compared with other PNs, \object{NGC\,7027} shows a low HNC abundance and an abnormal high N$_2$H$^+$ \citep{bachiller97}.
\citet{hasegawa01} suggested that most of the observed molecular lines are emitted from the PDR of the object except for CO emission which arises from a much more extended region.
%The observations of molecular lines provide a useful probe on the chemical and physical environments that they originate in. 
According to \citet{cox87}, lines from the ring molecule C$_3$H$_2$ are emitted from  dense regions. The observations by \citet{thronson86} and \citet{deguchi90} suggested that grain chemistry probably plays an important role for the formation of molecules in \object{NGC\,7027}.

The chemistry in the PDR of \object{NGC\,7027} has been modeled by 
\cite{yan99}, who suggested that the molecular abundance is significantly influenced by both photodissociation and shock chemistry.  A detailed chemical model of
the neutral envelope of \object{NGC\,7027} by \citet{hasegawa00} suggested that
most of the detected molecules in \object{NGC\,7027} are formed over a short timescale
after the object evolved into the PN stage. 
{ This was supported by the observations of \citet{herpin02}
 with the {\it Infrared Space Observatory (ISO)}. }
However, different conclusion was
obtained by \citet{josselin03} who suggested that most of molecules can survive during the transition from AGB to PN.
{ \citet{redman03} modeled the chemical evolution from
AGB to PPN and PN considering the affect of clumpiness. By comparing 
model results and observations, they proposed that the clumps form
at an early stage of PN and may be plentifully present within 
\object{NGC\,7027}.}

Hydrogen and helium recombination lines from \object{NGC\,7027}
are detectable at millimeter wavelengths,
allowing us to investigate the physical conditions and
dynamics of the ionized
regions. An advantage of the radio data is the reduced effect of dust 
extinction. A few radio recombination lines from
H56$\alpha$--H110$\alpha$ have been detected in \object{NGC\,7027}
\citep[][and the references therein]{roelfsema91,chaisson76,ershov89}.
\citet{ershov89} constructed a dynamical model to explain the
observed radio recombination lines and found that the expansion velocity
of the ionized gas is in good agreement with that of the molecular
envelope. \citet{roelfsema91} observed the
 H76$\alpha$ line observations and argued
that the ionized gas is decelerated by the surrounding molecular material.

In this paper, we report a systematic spectral survey of \object{NGC\,7027}
using the ARO 12\,m and the SMT 10\,m telescopes
\footnote{The 12\,m telescope and the
Heinrich Hertz Submillimeter Telescope (SMT) is operated by the Arizona Radio Observatory (ARO), Steward Observatory, University of Arizona.},
covering the frequency ranges 71--111\,GHz, 157--161\,GHz, and
218--267\,GHz. These data allow us to investigate the physical conditions
and chemical compositions of the molecular envelope
of this PN. This paper is organized as follows:
Sect.~2 describes the observations and the procedures of data reduction;
Sect.~3 presents the identification and measurements of detected
emission lines;  in Sect.~4 the method of data analysis is described
and the resulting excitation temperatures,
column densities and abundances relative to H$_2$ are presented;
in Sect.~5, we discuss the physical and chemical characteristics in the molecular envelope and ionized regions; and  the main conclusions are summarized in Sect.~6.

\section{Observations and data reduction}

The spectral survey was carried out between 2005 April and 2006 September. The observations were made in beam switching mode with an azimuth beam throw of 
{ 2\arcmin}. Pointing accuracy was checked every two hours.  The on-source integration time was more than one hour for each frequency setting.

The 71--111\,GHz and 157--161\,GHz spectra were obtained using
the ARO 12\,m telescope at Kitt Peak. The dual-channel
SIS receivers were employed in single sideband mode, yielding
a system temperature of 150--400\,K. The image rejection ratio 
was 16--20\,dB. The receiver back-ends were equipped with
two 256-channel filter banks (FBs) with spectral 
resolutions of 500\,kHz and 1\,MHz and a  millimeter autocorrelator 
(MAC) with 3072 channels and 195\,kHz resolution. 
All spectrometers were operated in series mode. 
{ Two orthogonal linear polarization modes were measured.}
The temperature
scale, $T^*_R$, was calibrated and corrected for atmospheric attenuation,
radiative loss, and rearward and forward scattering and spillover
by the chopper-wheel method. The main beam brightness
temperature was derived using
$T_R=T^*_R/\eta^*_m$, where $\eta^*_m$ is the corrected beam efficiency.
Over the frequency range 71--161\,GHz, $\eta^*_m$ is from 0.94 to 0.71,
{ and the conversion factor is 32.5--29.1\,Jy/K.}

The 218--267\,GHz observations were carried out with the
SMT 10\,m telescopes at Mt. Graham, Arizona, using 
the dual-channel SIS receivers operated in single sideband dual polarization
mode.
The system temperature is typically 400--700\,K.
The spectrometers used were a 2048-channel 
acousto-optical spectrometer (AOS) with a spectral resolution of
500\,kHz per channel and 1024-channel Forbes Filterbanks (FFBs)
with a spectral resolution of 1\,MHz per channel.
{ All spectrometers were used simultaneously.}
The data were calibrated to the antenna temperature scale, $T^*_A$, which
was corrected for atmospheric attenuation. $T^*_A$
is converted to main beam temperature by { $T_R=T^*_A/\eta_{mb}$}, where
the main beam efficiency, $\eta_{mb}$, is $\sim0.7$.
{ The conversion factor is 35\,Jy/K.}

All spectra were reduced using the CLASS software package in GILDAS
\footnote{GILDAS is developed and distributed by 
the Observatorie de Grenoble and IRAM.}.
%(see http://www.iram.fr/IRAMFR/GILDAS).}.
After discarding the bad scans which are seriously affected by bandpass
irregularities, we co-added the calibrated spectral data using
the $rms$ noise of each spectrum as weights.
A low-order polynomial baseline defined by the line-free spectral regions was subtracted from each spectrum.
{ In order to improve the signal-to-noise, the ARO 12\,m and SMT 10\,m 
spectra were smoothed and rebinned %by a factor of 3 
to a spectral resolution of 1\,MHz and 3\,MHz, respectively}, yielding 
a typical $rms$ noise temperature of 
$<8$\,mK in main beam temperature unit. 
%There are only a few spectral portions with a higher noise temperature because most of the scans  in those portions were seriously affected by bandpass irregularities and thus were discarded in obtaining the co-added spectra.
{
Based on a few strong lines detected in different spectrometers and
different epochs, we estimate that the calibration uncertainties introduced by
instruments and atmospheric conditions amount to about $15\%$.}

\section{The spectra}

\begin{figure*}
\vspace*{-12mm}
\plotone{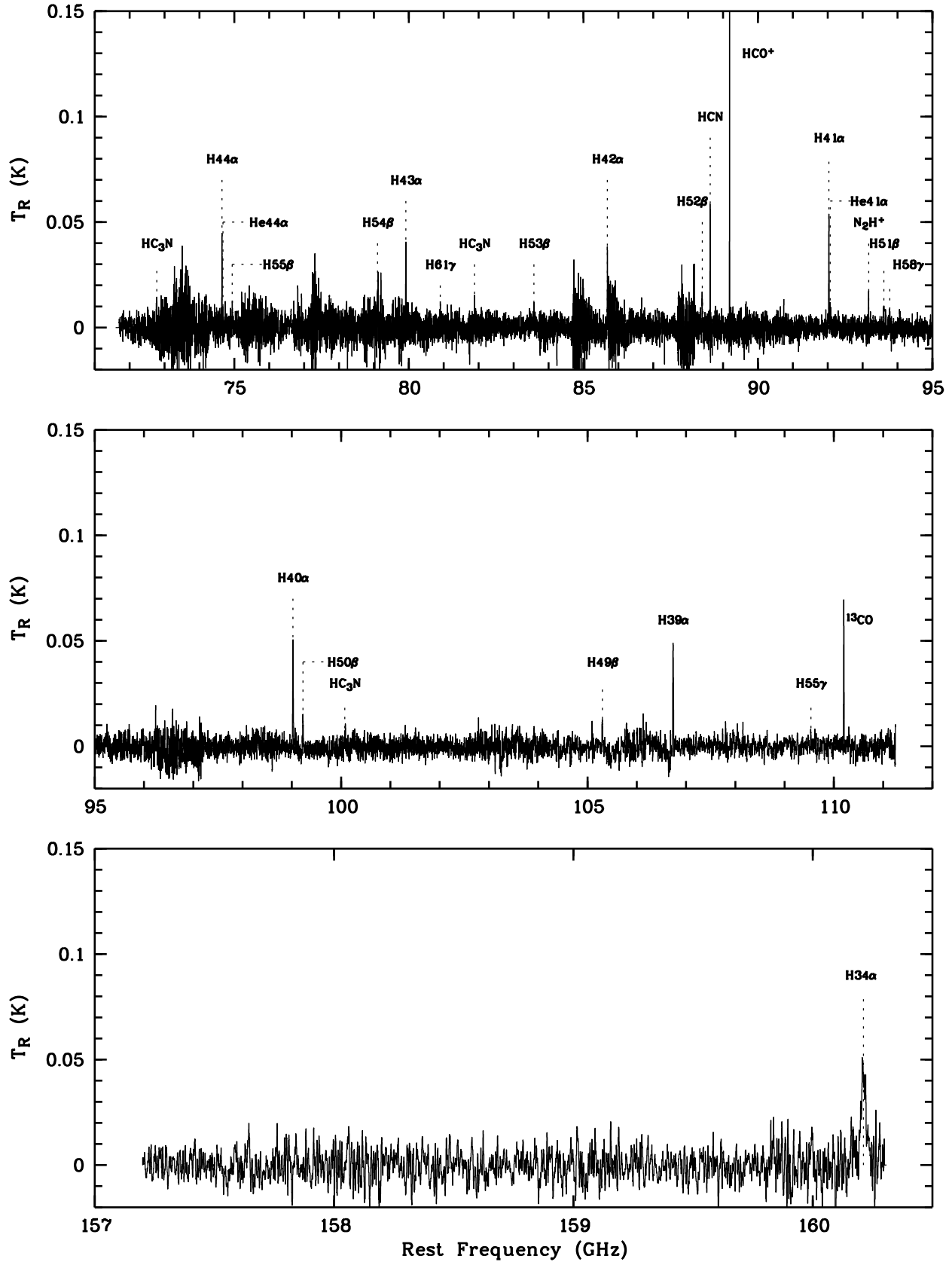}
\caption{From top to bottom: the 71--111\,GHz 157--161\,GHz and the  157--161\,GHz MAC spectra of NGC\,7027
obtained with the ARO 12\,m telescope. { The spectra have been smoothed to a resolution of 1\,MHz. The spectra at full resolution is available online in
 Fig.~\ref{spe_n7027_12m_ex}.}
}
\label{spe_n7027_12m}
\end{figure*}

\begin{figure*}
\plotone{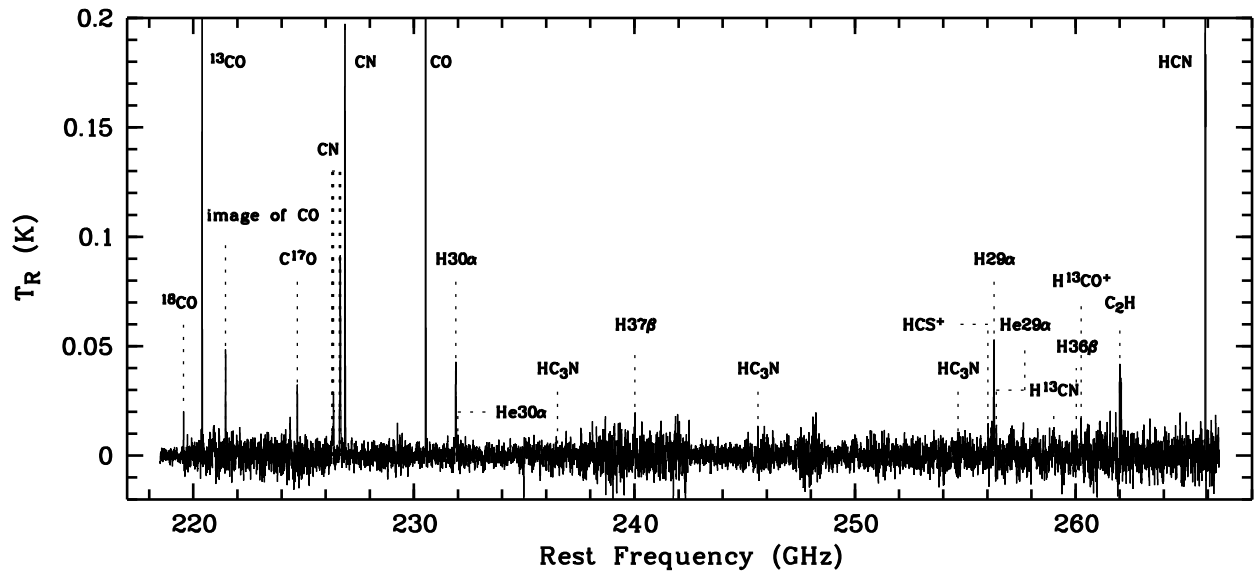}
\caption{The 218--267\,GHz FFB spectrum of NGC\,7027 obtained with the SMT 10\,m telescope.
{ The spectra have been smoothed to a  resolution of
3\,MHz. The spectra at full resolution is available online in
 Fig.~\ref{spe_n7027_smt_ex}.}  }
\label{spe_n7027_smt}
\end{figure*}

\begin{figure*}
\epsscale{.9}
\plotone{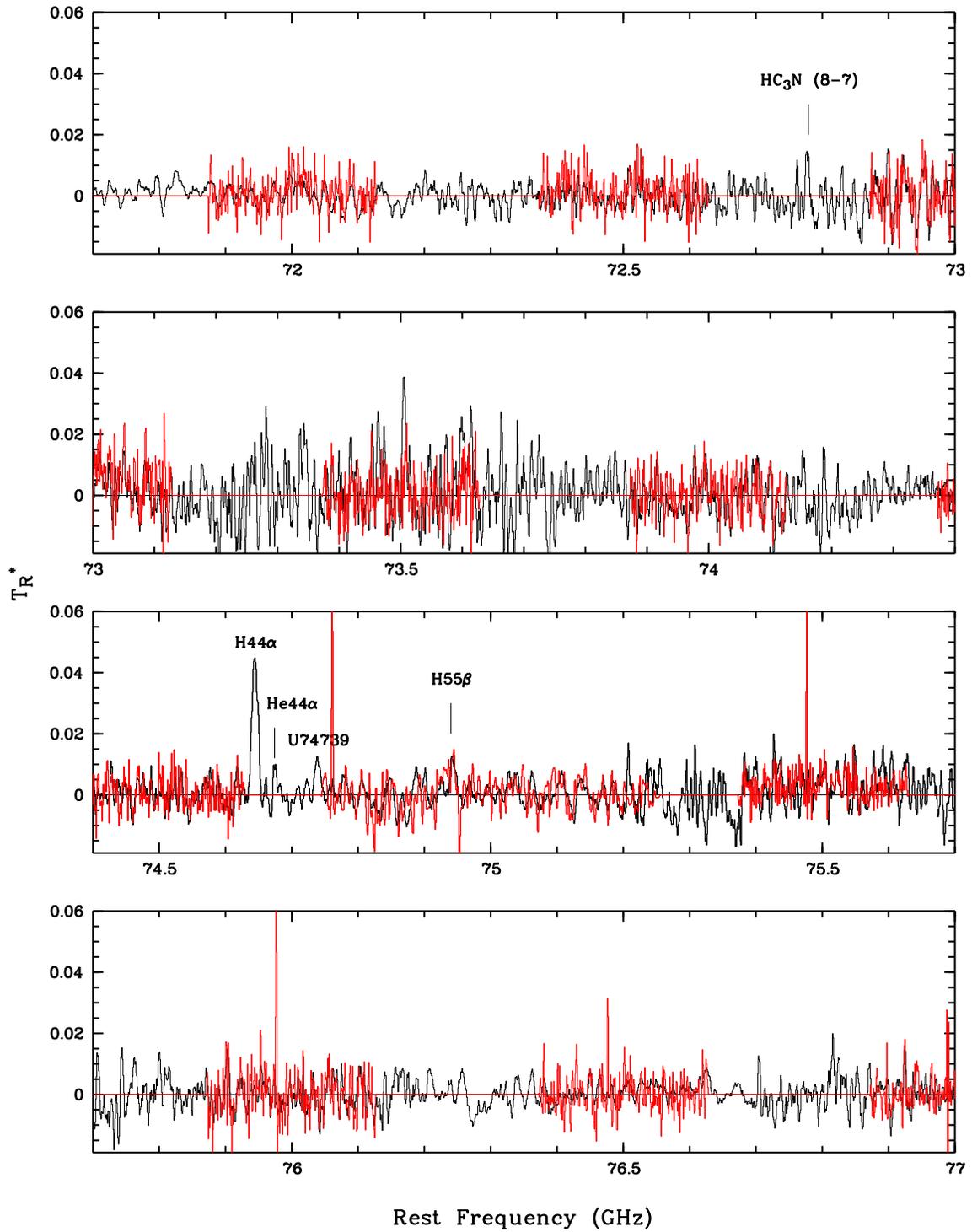}
\caption{The 71--111\,GHz and  157--161\,GHz spectra of NGC\,7027 obtained
with the ARO 12\,m
telescope. The black and red lines are the MAC and FB data, respectively.
The spectral resolution is 1\,MHz. The `:' represents uncertain detection.
 }
\label{spe_n7027_12m_ex}
\end{figure*}

\begin{figure*}
\plotone{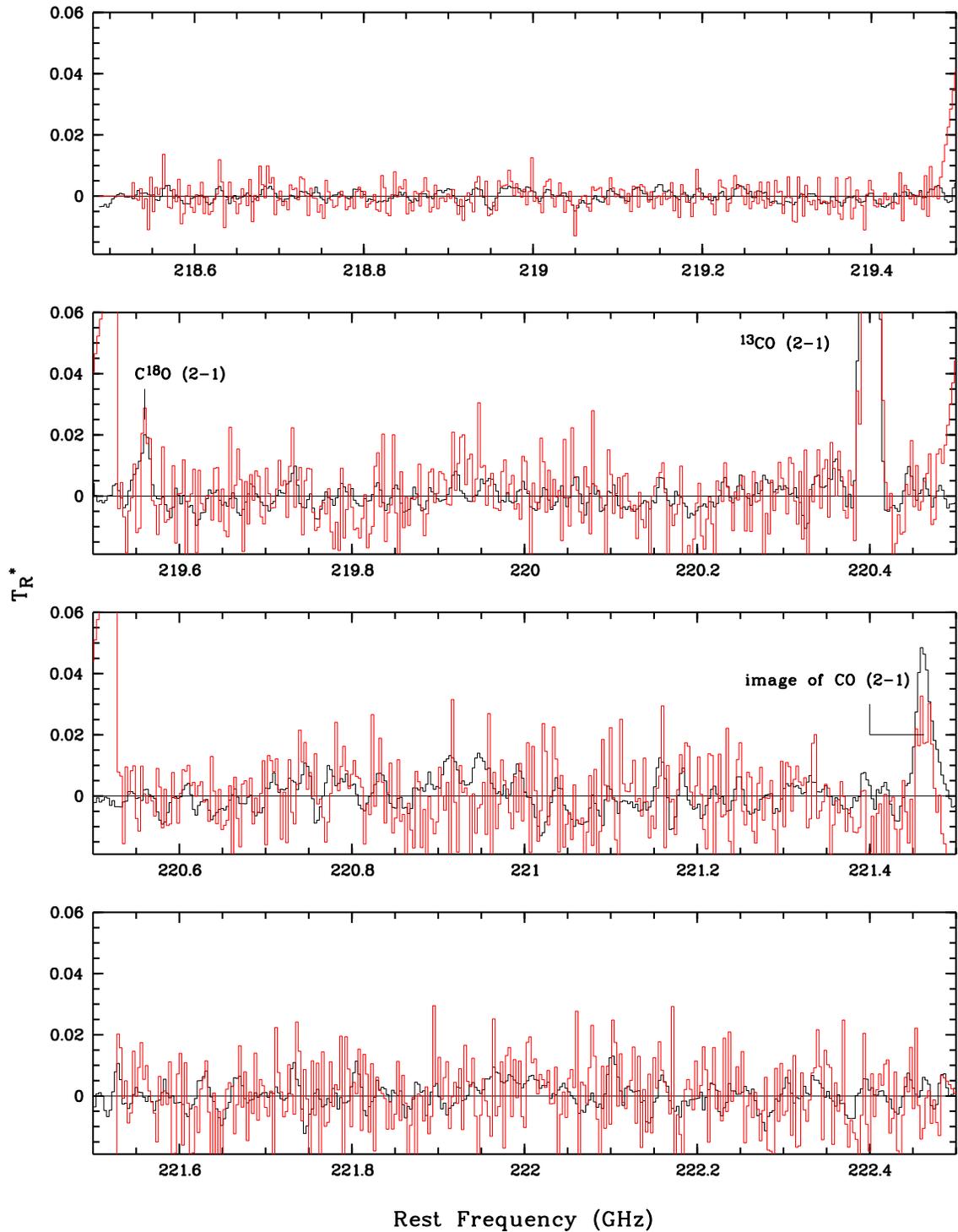}
\caption{The 218--267\,GHz spectrum of NGC\,7027 obtained with the SMT 10\,m
telescope. The black and red lines are the FFB and AOS data, respectively.
The spectral resolution is 3\,MHz. The `:' represents uncertain detection.
 }
\label{spe_n7027_smt_ex}
\end{figure*}

The spectra of \object{NGC\,7027} obtained with the ARO 12\,m telescope
and the SMT 10\,m telescope are presented in Fig.~\ref{spe_n7027_12m}
and Fig.~\ref{spe_n7027_smt}, respectively.  { The spectra at a full
resolution are shown in Fig.~\ref{spe_n7027_12m_ex} and 
Fig.~\ref{spe_n7027_smt_ex}, which is only available in the electronic 
version of this manuscript. Note that some features in these figures
are caused by bandpass irregularities.  More molecular lines were 
detected in the 10\,m spectra compared to the 12\,m spectra
as the ARO 12\,m telescope has a larger beam size and thus
bears larger beam dilution effect.}
Sections of the spectra containing spectral lines are expanded in Figs.~\ref{n7027_12m} (12\,m) and \ref{n7027_smt} (10\,m) in order to show the line profiles.   %A detailed discussion of the line profiles will be given below.
Since each spectrum was simultaneously observed with two receiver back-ends,
an emission line should be recorded by both spectrometers. This allows us to distinguish between real emission lines and artificial features.  
{  Contamination from image band was carefully checked.
The feature at 221.463\,GHz was found to be the CO (2--1) line 
from the image band.  No other lines from the image band were apparent.
A total of 71 distinct emission features were measured.}

The molecular lines are identified on the basis of the JPL catalog 
\citep{pickeet98}\footnote{http://spec.jpl.nasa.gov.} and the
Cologne database for molecular spectroscopy 
\citep[CDMS,][]{muller01,muller05}\footnote{http://www.ph1.uni-koeln.de/vorhersagen/.}.
The suggested identifications are labeled in Figs.~\ref{spe_n7027_12m}
and \ref{spe_n7027_smt}.
Combining the single and blended features, we identified 21 molecular lines, 
belonging to 8 molecular species (CO, CN, C$_2$H, HCN, HCO$^+$, HCS$^+$, 
HC$_3$N, N$_2$H$^+$) with a total of 13 isotopomers.
{ We also present `possible' detection of  C$_3$H$_2$,  HC$^{17}$O$^+$, 
and HC$^{18}$O$^+$. 14 lines remain unidentified.}
Several of these transitions are discovered for the first time in \object{NGC\,7027}.
%Compared to the low-frequency bands, more molecule lines were detected in the high-frequency bands.
A list of molecular lines detected in our spectra is presented in
Table~\ref{mol_n7027}. The table also gives the $rms$ noise, the main beam temperatures,
and the integrated intensities (in units of K km/s) of the lines.  
%With a few exceptions, most of the molecular lines have a parabolic profile. 
%The integrated intensities were obtained by direct integration over the observed line profile. Among the molecular lines discovered here, some have been observed by other authers, as given in the last column of Table~\ref{mol_n7027}.

The CO (2--1) and $^{13}$CO (1--0, 2--1) lines have been extensively  
observed by various telescopes in the past \citep[see, e.g.][]{hasegawa01}. The integrated intensities of the CO (2--1) and  $^{13}$CO (2--1)
lines presented in Table~\ref{mol_n7027}  are 
lower than those reported by \citet{hasegawa01} (see their Table~1). 
However, we note that
the CO(2--1)/$^{13}$CO(2--1) integrated intensity ratio of 24.8 is in excellent
agreement the value of 21.8 in \citet{hasegawa01}.
The lower antenna temperatures detected here are due to the larger beam dilution 
effect since the beam size of the SMT 10\,m telescope is a factor of $\sim2$ larger
than the 15\,m James Clarke Maxwell Telescope  used by \citet{hasegawa01}.
The C$^{18}$O (2--1) and  C$^{17}$O (2--1) lines were clearly detected in
our spectra. To our knowledge, it is the first time that these two
transitions are detected in \object{NGC\,7027}. 

Three strong CN (2--1) hyperfine groups are detected in our observations.  These lines 
have previously been seen by \citet{cox93}, \citet{josselin03}, and
 \citet{bachiller97}. The emission feature
at 227.2\,GHz is only marginally above our limit of detection and may be a combination of the CN ($N=2-1, J=5/2-1/2$) and the C$_3$H$_2$ (4$_{32}$--3$_{21}$) transitions. 
Other previous detections of CN in \object{NGC\,7027}
include the $J=3-2$ and $J=1-0$ transitions at 340 and
114\,GHz, respectively \citep{hasegawa01,bachiller97,thronson86,cox93,
josselin03}. 

The C$_2$H (3--2) transitions in \object{NGC\,7027},
first discovered by \citet{hasegawa01},  are confirmed in our observations.
All the C$_2$H (3--2) lines show a double-peaked profile.
The C$_3$H$_2$ (2$_{20}$--2$_{11}$) at 18.3\,GHz and the 
C$_3$H$_2$ (1$_{10}$--1$_{01}$) at 21.6\,GHz transitions were first detected
in the PN by \citet{cox87}. { We attempted to search for 
two new C$_3$H$_2$ transitions (6$_{25}$--5$_{14}$ and 5$_{33}$--4$_{22}$) at 
higher frequencies. The two lines are extremely faint, with intensities 
between 2--3$\sigma$ noise level. 
Hence, our detection of C$_3$H$_2$ should be 
treated with some caution.}

%We discovered two new
%C$_3$H$_2$ transitions (6$_{25}$--5$_{14}$ and 5$_{33}$--4$_{22}$) at 
%higher frequencies, confirming the presence of C$_3$H$_2$ in this PN. 

The well studied HCN (1--0, 3--2) and HCO$^+$ (3--2) transitions are also detected. We also detected their isotopic transitions H$^{13}$CN (3--2)
and H$^{13}$CO$^+$ (3--2). However, { the transitions
H$^{13}$CN (1--0) at 86.340\,GHz and H$^{13}$CO$^+$ (1--0) at
86.754\,GHz are overwhelmed by noise.
The transitions HC$^{17}$O$^+$ (1--0) and
HC$^{18}$O$^+$ (1--0) are only marginally above 2$\sigma$ noise level,
and thus have an ambiguous detection.}
A narrow feature at 256.028\,GHz has been discovered and identified with
the HCS$^+$ (6--5) transition, confirming the observations by
\citet{hasegawa01}. We also detect the N$_2$H$^+$ (1--0) line which was first reported by \citet{cox93}.

An unsuccessful search for the HC$_3$N (11--10) transition at 100.076\,GHz was made by \citet{thronson86}.\footnote{Note that in Table~1 of  \citet{thronson86}, the frequency of the HC$_3$N (11--10) transition
is given as ``110.0\,GHz'' by mistake.}
In our observations, several HC$_3$N lines are identified in the spectra 
from both the ARO 12\,m telescope and SMT 10\,m telescope.  
{ However, we do not detect the transitions HC$_3$N (10--9) at 90.979\,GHz
and HC$_3$N (12--11) at 109.174\,GHz, making our identification arguable.
If confirmed, this is the first detection of this species in a PN.}

%The unidentified line at 221.463\,GHz is strong and has a well-defined profile.  We  cannot find any mention of this line in the literature in \object{NGC\,7027} or other objects.
{ The unidentified lines are listed in Table~\ref{un_n7027}. Most of them are very faint, and
are only marginally detected. Some of these lines have been detected in the star formation regions
Sgr B2(N) and Sgr B2(M) \citep{tur89,num98}. We also compared our spectra with a recent
molecular line survey of the PPN \object{CRL\,618} by \citet{pardo07}.
These U-lines are not detected in their spectra.
Thus we infer that some of these lines probably arise from species which are formed in very strong UV 
radiation. }

{
Table~\ref{limit} lists a number of molecular species which are not 
detected by our observations of \object{NGC\,7027}.  For a given
species, the strongest transition in our frequency ranges is given in
the table aiming to reach the tightest constrain on its column density.}  
The upper limits of the integrated intensities can be
obtained by $3\sigma \Delta v/\sqrt{n}$, where $\sigma$ is the $rms$ noise, and $n$ is the number of { channels within the line width $\Delta v$.}

A number of recombination lines, including 24 hydrogen lines and
8 helium lines, are detected in our observations. Previously, most of the
recombination lines were observed in the low-frequency bands.
The identifications of recombination lines are based on the calculations by \citet{lilley68} and 
\citet{towle96}. No heavy-element recombination line is visible in
our spectra as they are under our detection limit. The identifications and 
measurements of the recombination lines are presented in Table~\ref{re_n7027}.
All recombination lines were measured using Gaussian line profile fitting.
Some of the lines, such as H36$\beta$, are probably  blended with unknown
features and show a relatively broad profile. The H60$\gamma$ line 
at 84.914\,GHz The H60$\gamma$ line and The He42$\alpha$ line at 85.723\,GHz
fall within the spectral regions which have a low signal-to-noise
(see Fig.~\ref{spe_n7027_12m}), and thus their measurements are uncertain.

\section{Analysis}

\subsection{Excitation temperatures and column densities}

In order to obtain molecular excitation temperatures and column densities,
the effect of beam dilution is corrected under
the assumption that both the source brightness
distribution and the antenna beam have a Gaussian profile.
The source brightness temperature (which is equivalent to source intensity) was obtained
by $T_s=T_R(\theta^2_b+\theta^2_s)/\theta^2_s$, where the main-beam brightness temperature $T_R$ has the physical meaning of average intensity, $\theta_b$ is
the antenna full beam at half-power and $\theta_s$ is the source diameter.
The beam sizes of the ARO 12\,m telescope and the SMT 10\,m telescope
are 86$''$--38$''$ and 32$''$--28$''$ over the frequency ranges 71--161\,GHz
and  218--268\,GHz, respectively. Different molecular species in
\object{NGC\,7027} might have different distribution and $\theta_s$ thus
should be separately estimated in terms of map of each species.
Based on the observations of \citet{hasegawa01}, we have assumed $\theta_s=13\farcs2$ for all the species
except CO. High-resolution mapping observations of CO in \object{NGC\,7027} suggest that the CO emission originates in a more extended region \citep{masson85}, so we adopted $\theta_s=40\arcsec$ for CO and its isotopomers.

The standard rotation-diagram method was applied to determine molecular
excitation temperatures and column densities. Under the assumption that
i) the lines are optically thin, ii) the level populations are in LTE and
can be described by a Boltzmann temperature, referred to as the 
``excitation  temperature'' ($T_{ex}$), and
iii) $T_{ex}$ is much larger than the background 
temperature ($T_{bg}$), we have the  relation

\begin{equation}
\ln \frac{N_u}{g_u}=\ln\frac{3k\int T_s dv}{8\pi^3\nu S\mu^2}=
\ln\frac{N}{Q(T_{ex})}-\frac{E_u}{kT_{ex}},
\label{boltzmann}
\end{equation}
where $N_u$, $g_u$, and  $E_u$ is
the population, degeneracy, and excitation energy of the upper level,
$\int T_s dv$ is the integrated intensity of the source, $S$ the line strength,
$\mu$ the dipole moment,
$\nu$ the line frequency,
 $Q$ the rotation 
partition function, and $N$ the column density.
If many transitions for an individual molecule
are observed, a straight line can be fitted by plotting the
integrated intensity versus upper level energy. Then
$T_{ex}$ and the total column density can be deduced from the slope 
($-1/T_{ex}$) and intercept [$\ln(N/Q)$] of this line.

{
Albeit with some problems, we probably detect five HC$_3$N transitions.  
The excitation energies of the upper levels cover a large range with 10\,K$<E/k<180$\,K, allowing us to determine excitation temperature and column density.
Fig.~\ref{dia_n7027} shows the rotation diagram for HC$_3$N. A linear least-squares fit to all the transitions produces $N=4.39\times10^{13}$\,cm$^{-2}$.
However, the data in the rotation diagram suggest that the low $E_u$ 
transitions seem to arise from a colder region.
 The results presented in \citet{hasegawa00} suggest that the neutral
envelope of \object{NGC\,7027} consists of a geometrically thin and dense shell and an extended outer stellar wind region, with the temperature in the dense shell being more than a factor of 10 higher
than that in the wind region. In this picture, our HC$_3$N  high $E_u$ transitions would originate dominantly in the hot, dense shell, whereas the cold stellar wind region contributes significantly to the excitation of low $E_u$ lines. 
The stratified temperature structure may cause some problems in the calculations
of column density. To get more precise results, detailed modelling is 
required, as done by \citet{pardo04} and \citet{pardo07b} for \object{CRL\,618}.
However, this is hampered by the weakness of most molecular
lines in \object{NGC\,7027}.
}

No rotation diagram is presented for other molecular species because
either only one line for the species was detected, or the upper levels of the observed transitions have similar excitation energies. 
We thus adopted a constant excitation temperature of
$T_{ex}=34.6$\,K, as deduced from the HC$_3$N rotation diagram, 
and applied Equ.~(\ref{boltzmann}) to estimate the column densities of
CN, C$_2$H, C$_3$H$_2$, HCS$^+$, N$_2$H$^+$, and HCO$^+$.
We derived a column density of $N=1.02\times10^{14}$\,cm$^{-2}$ for CN,
which is consistent with the values of $3.5\times10^{14}$\,cm$^{-2}$ and
$6.1\times10^{13}$\,cm$^{-2}$ deduced by
\citet{hasegawa01} and \citet{josselin03}, respectively.
Our analysis yields $N=3.49\times10^{14}$\,cm$^{-2}$ for C$_2$H, in
good agreement with the value of
$2.4\times10^{14}$\,cm$^{-2}$ given by \citet{hasegawa00}. From the
observations of the C$_3$H$_2$($1_{10}$--$1_{01}$) at 18.3\,GHz
C$_3$H$_2$($2_{20}$--$2_{11}$) at 21.6\,GHz transitions,
\citet{cox87} found $N$(C$_3$H$_2$)$\sim3\times10^{12}$\,cm$^{-2}$,
which is lower than our value ($1.11\times10^{14}$\,cm$^{-2}$) by a factor of 
about 36. { Given large measurement errors, our value may be unreliable.}
For HCS$^+$, we obtain a value of $N=4.96\times10^{12}$\,cm$^{-2}$, which should be  treated with some caution as it was derived from an extremely faint emission line.
From the  N$_2$H$^+$(1--0) line at 93.2\,GHz, we derive $N$(N$_2$H$^+$)$=2.43\times10^{13}$\,cm$^{-2}$.
The column densities of HCS$^+$ and N$_2$H$^+$ have not been reported previously.
{ Our determination of
$N$(HCO$^+$) is higher than that of
\citet{hasegawa01} by a factor of 5.
Considering the uncertainties in calibrations, such difference is not unreasonable.
}

The strong HCN and CO transitions are likely to be optically thick and therefore the rotation diagram analysis is not applicable.   For these lines we used the following equations to deduce their excitation temperatures and column densities. 
%and their isotopomers of HCN, HCO$^+$, and CO have also been observed.

The equation of radiative transfer is
\begin{equation}
T_s=[I(T_{ex})-I(T_{bg})](1-e^{-\tau}),
\label{transfer}
\end{equation}
where $I(T)=h\nu[k(e^{h\nu/kT}-1)]^{-1}$ with
$h$ the Planck constant and $k$ the Boltzmann constant, 
$\tau$ is the optical depth at the center of a line, and $T_{bg}$ is the cosmic background
temperature (2.7\,K). For the optically thick lines ($\tau\gg1$), $T_{ex}$ can be obtained from Equ.~(\ref{transfer}) and is given by

\begin{equation}\label{ts}
T_{ex}=\frac{h\nu}{k\ln\left\{\frac{hv}{k[T_s+I(T_{bg})]}+1\right\}}.
\end{equation}

Assuming that the isotopomer has the same excitation temperature as the
main line, we can obtain its optical depth at the line center from Equ.~(\ref{transfer})
\begin{equation}
\tau_i=\ln\left[1-\frac{T_s}{I(T_{ex})-I(T_{bg})} \right].
\label{tau}
\end{equation}

The column density of the isotopomer can be derived from

\begin{equation}
N_i=\frac{3h\Delta v\tau_iQ(T_{ex})\exp(\frac{E_u}{kT_{ex}})}{8\pi^3\mu^2S
[\exp(\frac{h\nu}{kT_{ex}})-1]},
\end{equation}
where $\Delta v$ is the line width.

Assuming that lines of the isotopomer is optically thin, the $T_s$ ratio of the main line and
the isotopic line can be derived using Equ.~(\ref{ts}):

\begin{equation}\label{col}
\frac{T_{s,m}}{T_{s,i}}\approx\frac{1-\exp(-\tau_m)}{\tau_i}.
\end{equation}
Once $\tau_i$ is obtained from Equ.~(\ref{tau}), the optical depth of the main line, $\tau_m$, 
can be obtained from the above equation.

Finally, we can determine the column density of the main species
using
\begin{equation}
\frac{N_m}{N_i}\approx\frac{\tau_m}{\tau_i}.
\end{equation}

%Applying the above equations to HCN, HCO$^+$, and CO, we obtain
%the optical depths at the line center,
%$\tau$(HCN)$=1.1$, $\tau$(H$^{13}$CN)$=0.057$, $\tau$(HCO$^+$)$=4.8$,
%$\tau$(H$^{13}$CO$^+$)$=0.036$,  $\tau$(H$^{18}$CO$^+$)$=0.072$,
%$\tau$(CO)$=2.7$, $\tau$($^{13}$CO)$=0.037$, 
%$\tau$(C$^{18}$O)$=0.004$, and $\tau$(C$^{17}$O)$=0.007$.

Applying the above equations to HCN and CO,
we obtained their excitation temperatures and column densities.
The results are summarized in Table~\ref{col_n7027}. 
{ Our determination of the column density of HCN is a factor of 16  
higher than that derived
by \citet{hasegawa01}. In our calculations,
a large error may be caused by the very faint
H$^{13}$CN line. If applying Equ.~(\ref{boltzmann}), we derive a lower limit
of $N$(HCN)$=4.2\times10^{13}$\,cm$^{-2}$, in agreement with the
result of \citet{hasegawa01}. }

For comparison,
Table~\ref{col_n7027} also lists the column densities predicted by
the model of \citet{hasegawa00}. The column densities deduced by
current work are in of the same approximate orders of magnitude with the model predictions except for
CO and N$_2$H$^+$. The model overestimation of the column density of CO is addressed by \citet{hasegawa00}, and an improved model \citep{hasegawa01} gives a better fit to the observed column 
density of CO.  The model, however, fails to explain the extremely high column density of N$_2$H$^+$.
A detailed discussion will be presented in the next section.

Assuming a line width of 30\,km s$^{-1}$, we estimate the upper limits of the column densities for several non-detected molecular species,  and these values are presented in Table~\ref{limit}.
Since these upper limits may still serve as useful constraints of  the chemical processes in \object{NGC\,7027}, these upper limits are compared with the model predictions of \citet{hasegawa00}.  Table~\ref{limit} shows that the observed upper limits are in reasonable agreement with the model predictions except for CS, which observed column density is lower than the model prediction. 
Even in the  improved model of \citet{hasegawa01}, the theoretical column density of $N({\rm CS})=8.4\times12$\,cm$^{-2}$ is still higher than the upper limit derived in the present study.

\subsection{Abundances calculations}

In order to calculate the fractional abundance of a molecular species
relative to H$_2$,
we need to know the number density of the species and 
that of H$_2$, which can be derived
by solving the coupled radiative transfer and statistical
equilibrium equations.
A full treatment of the line excitation requires knowledge of the 
geometrical structure of the envelope, the
rates for collisional excitation of the molecules,
and the effects of dust emission and absorption, which however is
poorly known, and is beyond the scope of this paper.
Instead, assuming that all molecular emission originate in a spherical
shell, we calculated the molecular abundances respect to H$_2$ using
the expression suggested by \citet{olofsson96},
\begin{equation}\label{abundance}
f_{\rm X}=1.7\times10^{-28}\frac{v_e\theta_bD}{\dot{M}_{{\rm H}_2}}
\frac{Q(T_{ex})\nu_{ul}^2}{g_uA_{ul}}
\frac{e^{E_l/kT_{ex}}\int T_Rdv}{\int^{x_e}_{x_i}e^{-4\ln2x^2}dx},
\end{equation}
where the integrated intensity $\int T_Rdv$ is given in K\,km\,s$^{-1}$,
the full half power beam width $\theta_b$ is in arc\,sec,
$v_e$ is the expansion velocity given in km\,s$^{-1}$, $D$
is the distance in pc, $\dot{M}_{{\rm H}_2}$ is
the mass loss rate in $M_{\sun}\,{\rm yr}^{-1}$, $\nu_{ul}$  the
line frequency in GHz, $g_u$ the statistical weight of
the upper level, $A_{ul}$ the Einstein coefficient for the transition,
$E_l$ the energy of the lower level, and $x_{i,e}=R_{i,e}/(\theta_bD)$
with $R_i$ and $R_e$ the inner radius and outer radius of the shell.
The equation is deduced under the assumption that
the lines are optically thin, $T_{ex}$ is uniform throughout the
shell, the molecular density follows an $r^{-2}$ law,
and the shell is formed by a constant mass loss rate and has
a constant expansion velocity. One should bear in mind that
the spatial variations in the fractional abundances and excitations 
may be a large source of error. For the calculations, we adopted
the excitation temperatures deduced in section~4.1 (see Table~\ref{col_n7027})
and a distance 880\,pc to \object{NGC\,7027} \citep{masson89}.
Following the model of \citet{hasegawa00},
we assumed $R_i=0.017$\,pc and $R_e=0.14$\,pc. The expansion velocity
was determined from the widths of a few strong lines, which suggest
$v_e\sim30$\,km\,s$^{-1}$.

Assuming that the CO (2--1) emission arises from the stellar-wind region, we can estimate the mass loss rate from the line strength using the formula given by
\citet{winters02},
\begin{equation}
\dot{M}=5.7\times10^{-20}\frac{T_{R}v^2_eD^2\theta_b^2}{S(J)f^{0.85}_{\rm CO}} M_{\sun}\,{\rm yr}^{-1},
\end{equation}
where the correction factor $S(J)=0.5$ for the CO (2--1) line,
the distance $D=880$\,pc, the abundance of CO relative 
to H$_2$, $f_{\rm CO}$, is assumed to be
$1\times10^{-3}$, and the expansion velocity $v_e=30$\,km\,s$^{-1}$.  The derived mass loss rate of $1.1\times10^{-4}$\,$M_\sun {\rm yr}^{-1}$ is in good accord with the value of $1.4\times10^{-4}$\,$M_\sun\,{\rm yr}^{-1}$ derived from the model calculations by \citet{sopka89}. The error in the determination of
$\dot{M}$ due to the distance uncertainty is $34\%$.
%The higher mass loss rate in \object{NGC\,7027} compared to AGB stars \citep[$10^{-6}$--$10^{-5}$\,$M_\sun {\rm yr}^{-1}$;][]{woods03} is a result of the presence of supperwind during the post-AGB phase \citep[see][for a detailed study]{herman85}.

Using Equ.~(\ref{abundance}), we derived $f_{\rm X}$ for all the observed
molecular species and the results are given in Table~\ref{col_n7027}. 
The calculated abundance ratios between different species are not 
rigorously equal to the ratios of column densities. This is due to two 
reasons.
One is the different assumptions made to deconvolve the brightness
temperature distribution of the source from the antenna beam.
For the calculations of column densities,
we have simply assumed that the source brightness distribution is Gaussian.
Second, for the abundance calculations, we have assumed that the lines 
are optically thin. Consequently, when the emission is optically thick, 
$f_{\rm X}$ given in the table should be treated as a lower limit.
We can use the $N({\rm X})/f_{\rm X}$ ratio, which approximately
gives the column density of H$_2$, to estimate
the error due to these assumptions. The last column of Table~\ref{col_n7027}
gives the $N({\rm X})/f_{\rm X}$ ratio deduced from each molecular species.
These values
are generally in good agreement except that derived from CO, which is
outside the 3$\sigma$ range. The high $N({\rm CO})/f_{\rm CO}$ ratio
gives a upper limit of $N$(H$_2$) since the CO emission is certainly 
optically thick. Excluding the value deduced from CO,
we obtained the average H$_2$ column density of
N$({\rm H}_2)=(7.7\pm3.8)\times10^{21}$\,cm$^{-2}$. Hence,
the error due to the assumptions of structure and optically-thin emission
should be less than a factor of 2. Combining with the uncertainties caused by
various parameters ({ excitation temperature}, distance, source size 
etc.) and our measurements,
we crudely estimate that the total error in the calculations of column 
density and abundance is a factor of $\sim8$. The column density ratios
and the abundance ratios are expected to be more reliable.

\section{Results and Discussion}

\subsection{Chemistry}

\subsubsection{{\rm CN} and {\rm HCN}}

Given their relatively large dipole moments, the rotational transitions from the molecules
CN and HCN are often detected in molecular clouds.
HCN can be formed through the dissociative recombination of
HCNH$^+$, while CN is dominantly produced via the photodissociation
of HCN by UV radiation. { Therefore, the CN abundance dramatically 
increases  with increasing stellar radiation when a star evolves beyond the
AGB, through the proto-PN  stage and to the PN stage.} Such a trend has been revealed
by a sample study of \citet{bachiller97}. 
{ There is observational evidence showing that the abundance of HCN
increases in the PPN stage, and then dramatically decreases in the PN
stage \citep{herpin02}. This can be attributed to different roles that
dissociative recombination and photodissociation play
in the two stages.} Our observation
yield a HCN/$^{13}$CO abundance ratios of $>0.008$, consistent 
the value of $>0.005$ reported by \citet{bachiller97}.
However, we do not find large CN enhancement in \object{NGC\,7027}.
The CN/H$_2$ abundance ratio is $7.0\times10^{-8}$,
about a factor of 10  lower than those in AGB stars reported by \citet{woods03}.
The model of \citet{hasegawa00} seems to overestimate
the CN abundance by a factor of 2--3 although this may be explained with
the uncertainties in the measurements and calculations. The unexpectedly
low CN abundance was also found by \citet{ali01} in their modelling of general 
PNs.

It has been generally accepted that the CN/HCN abundance ratio
can trace the UV radiation field.
\citet{bachiller97} suggest that after a star evolves into
the PN stage, the CN/HCN abundance ratio remains constant
($\sim9$).
We obtain $N$(CN)/$N$(HCN)$<2.4$ in \object{NGC\,7027}.
This probably suggests that the  reaction
\begin{equation}\label{hcn}
{\rm CN+H_2 \rightarrow HCN+H}
\end{equation}
is very efficient  in this young PN.

\subsubsection{\rm HNC}

HNC can be formed in a similar way as HCN.
{ \citet{herpin02} suggested that in PPN stage
the HNC abundance is 
enhanced with respect to HCN through ion-molecule reactions.}
Both observations \citep{sopka89,bachiller97,josselin03} and models \citep{ali01} 
indicate that the average abundance ratio HNC/HCN in PNs is
$\sim 0.5$. 
However, we do not detect HNC emission in \object{NGC\,7027}. The
HCN/HNC ratio is estimated to be larger than 30.
 The abnormally low HNC abundance
in \object{NGC\,7027} was also noted by \citet{bachiller97} and 
\citet{josselin03}, who obtained $N({\rm HCN})/N({\rm HNC})>17$. 
As proposed by
\citet{bachiller97}, HNC may have been destroyed in \object{NGC\,7027}
through 
\begin{equation}
{\rm HNC+H\rightarrow HCN+H},
\end{equation}
which is efficient in high temperatures. The hypothesis is supported by the
high HCN abundance.
The model of \citet{hasegawa00} did not include the HNC reaction chains.
A more comprehensive model is required
to account for the observed CN/HCN/HNC abundance ratio.

\subsubsection{\rm HC$_3$N}

HC$_3$N is commonly observed in AGB stars and proto-PNs
{ \citep[e.g.][]{cernicharo00,pardo04}. 
 \citet{pardo05} found that HC$_3$N is quickly reprocessed
from HCN in PPN stage.}
Prior to this work, HC$_3$N
has never been detected in PNs and its absence is often explained as the result of efficient photodisociation due to the strong UV radiation in PNs.
\citet{thronson86} estimated
the abundance ratio $N({\rm HC_3N})/N({\rm CO})\le1\times10^{-4}$ in \object{NGC\,7027}. 
The abundance upper limit of HC$_3$N is comparable with that in 
\object{IRC+10216}. They thus did not find evidence for underabundance
of HC$_3$N in the PN.

{ Here we report a possible detection of HC$_3$N in the PN. 
Our results yield $N({\rm HC_3N})/N({\rm CO})=1.2\times10^{-5}$, }
at lest one order of magnitude lower than that in \object{IRC+10216},
confirming the destruction caused by UV photons.
The time-dependent chemical model of \citet{ali01} predicts
that the HC$_3$N abundance gradually decreases with the
evolution of PNs. Their model predicted that \object{NGC\,6781},
a more evolved PN, has a
$N({\rm HC_3N})/N({\rm CO})$ ratio of $4.2\times10^{-9}$, which is three orders of magnitudes lower than
that in \object{NGC\,7027}.
%It is impossible that the coefficient of photodissociation dramatically increases during the early stages of PN evolution.
The higher abundance of HC$_3$N derived here may be the result of the presence of dense clumps  within the  nebula, which serve to shield the molecules from the UV radiation from the central star.

\subsubsection{\rm C$_2$H}

The C$_2$H abundance in 
\object{NGC\,7027} is about a factor of 50 lower than that in
\object{IRC+10216}. Based on a comparison between the
observation and the model prediction of \citet{hasegawa00},
\citet{hasegawa01} concluded that C$_2$H  is newly produced via the
photodissociation  of C$_2$H$_2$ instead of being left over from AGB phase. The C$_2$H abundance derived in the present study  is in good agreement with the value given by their observation and model.

\subsubsection{\rm C$_3$H$_2$}

The C$_3$H$_2$ abundance in
\object{NGC\,7027} is { at least} a factor of 4 lower than that in
\object{IRC+10216}. 
Albeit with a low abundance, the
large dipole moment makes C$_3$H$_2$ detectable. 
{ Although our detection of C$_3$H$_2$ is uncertain,
the observation of \citet{cox87} showed the presence of this species.}
The abundance of C$_3$H$_2$ was not predicted by
the model of \citet{hasegawa00}. 
There is no
path for the formation of  C$_3$H$_2$ in PNs, and thus it may be
a remnant from AGB envelope.
Based on the observation of molecular clouds,
\citet{turner91} found that  C$_3$H$_2$ avoids the hot environment
and favors the cool clouds. Consequently,
to explain its survival from the dissociation of strong UV
radiation field in \object{NGC\,7027},
we infer that C$_3$H$_2$ may arise from some
cool high-density clumps.

\subsubsection{{\rm HCO$^+$} and {\rm N$_2$H$^+$}}

HCO$^+$ and N$_2$H$^+$ can be formed through reactions,
\begin{equation}\label{hco}
{\rm
H_3^++CO\rightarrow HCO^++H_2,
}
\end{equation} 
and
\begin{equation}\label{hn}
{\rm H_3^++N_2\rightarrow N_2H^++H_2,} 
\end{equation}
which are expected to have a similar rate coefficient.
Both HCO$^+$ and N$_2$H$^+$ can be destroyed by dissociative electron 
recombination.
Therefore, if reactions~(\ref{hco}) and (\ref{hn}) dominate the
production of HCO$^+$ and N$_2$H$^+$, we have the abundance
ratio ${\rm HCO^+/N_2H^+\approx CO/N_2}$. However, HCO$^+$ 
can be produced through additional processes, such as
\begin{equation} \label{hco1}
{\rm CO^++H_2\rightarrow HCO^++H,}
\end{equation}
while N$_2$H$^+$ has no alternative formation channel and can be destroyed 
at high density by proton transfer reaction with CO,
\begin{equation}\label{hn1}
{\rm
N_2H^++CO\rightarrow HCO^++N_2.
}
\end{equation}
As a result, the ${\rm HCO^+/N_2H^+}$ ratio is expected 
to be larger than the CO/N$_2$ ratio. 

The observations of HCO$^+$ and N$_2$H$^+$ allow us to
test these ion-molecule reactions in \object{NGC\,7027}.
Assuming that the N$_2$/H$_2$ abundance ratio is equal to N/H
\citep[1.4$\times10^{-4}$,][]{zhang05} and $N({\rm CO})/N({\rm H_2})=1\times10^{-3}$,
we have $N({\rm CO})/N({\rm N_2})=7.1$.  On the other hand, our observations
yield
$N({\rm HCO^+})/N({\rm N_2H^+})=14$, 
higher than the CO/N$_2$ abundance ratio
by a factor of 2. Hence, additional formation processes for HCO$^+$
may be significant.

The N$_2$H$^+$ abundance in \object{NGC\,7027} is abnormally high, with an observed column density 7--8 orders of magnitude higher than predicted by the model of \citet{hasegawa00}. The species, however, has 
not been detected in more evolved PNs \citep{bachiller97}. Since reaction (\ref{hn})
is the unique path for the production of N$_2$H$^+$, the high N$_2$H$^+$ abundance therefore implies a high formation rate of H$_3^+$.
H$_3^+$ can be produced through ionization of H$_2$ by
cosmic ray or soft X-ray emission from the central star. Since strong
X-ray emission  from \object{NGC\,7027} has been detected 
\citep{kastner01}, it is possible  that X-ray emission may
play an key role in the formation of N$_2$H$^+$ in \object{NGC\,7027}.
This is consistent with the fact that more evolved PNs generally have weaker X-ray emission and have no N$_2$H$^+$ emission. If high N$_2$H$^+$ abundance is a property of young PNs, further observations of N$_2$H$^+$ in a sample of young PNs would be useful.

Although N$_2$D$^+$ has been extensively observed in dark clouds, we are unable to detect it in \object{NGC\,7027}.   From the 3$\sigma$ intensity upper limit of the 
N$_2$D$^+$(1--0) transition, we estimate that  \object{NGC\,7027}
has a fractionation ratio $N({\rm N}_2{\rm D}^+)/N({\rm N}_2{\rm H}^+)<0.1$.
The lower fractionation ratio in \object{NGC\,7027} compared to that
found in dark clouds \citep{daniel07} can be attributed to
different physical environments.

HCO$^+$ in \object{NGC\,7027} has a similarly high abundance with
that in older PNs.
The HCO$^+$/HCN abundance ratio in \object{NGC\,7027} is 0.5, which is in very good
agreement with the average value for older PNs \citep{bachiller97}.
The different behaviors of HCO$^+$ and N$_2$H$^+$
in young and more evolved PNs suggest that 
reaction~(\ref{hco}) is not the dominant process for the formation of HCO$^+$.

\subsubsection{{\rm HCS$^+$} and {\rm CS}}

We obtain a $N({\rm HCO^+})/N({\rm HCS^+})$ column density ratio of 68, which is comparable to
the ion abundance ratio of $N({\rm O}^+)/N({\rm S}^+)=111$ derived in the ionized regions
of \object{NGC\,7027} \citep{zhang05}.   Hence, the 
formation rate of HCS$^+$ should be similar with that of HCO$^+$.

%The upper limit of the CS/HCN ratio in \object{NGC\,7027} is about 30 times
%lower than the CS/HCN ratio in \object{IRC+10216}, and 
The estimated CS abundance is about a factor of 10 lower than the model  value of 53 \citep{hasegawa00}.
% Moreover, the CN/CS ratio is 53, which is about 3 times larger than the N/S abundance ratio \citep[$\sim$17,][]{zhang05}. 
{ According to \citet{herpin02},  after a star leaves the AGB strong shocks can occur and significantly affect the circumstellar chemistry.}
\citet{willacy98} suggest that shocks can destroy CS and HCN in circumstellar
envelopes, which might imply that the low CS abundance observed in \object{NGC\,7027} is due to the presence of shocks.
%However, we do not find evidence for underabundance of HCN in 
%\object{NGC\,7027}, suggesting that the shock destruction of HCN may be compensated
%by reaction~(\ref{hcn}).

\subsection{Isotopic ratios}

\subsubsection{Carbon}

The determination of isotopic abundances plays an important role in our understanding of 
nucleosynthesis in evolved stars. Extensive studies have shown
that the $^{12}$C/$^{13}$C abundance ratios in red giant
stars are considerably lower than the prediction made by standard
stellar evolution models \citep[e.g.][]{charbonnel98}.
An extra mixing mechanism, called cool bottom processing (CBP),
was introduced to explain the low $^{12}$C/$^{13}$C ratio
\citep{sackmann99,boothroyd99}.
{ For the stars with masses in the range $2.5\le M/M_\sun \le 6$,}
 the hot bottom burning (HBB) may occur during
the AGB stage and induce $^{12}$C/$^{13}$C to further decrease to 
$\sim3.5$ \citep{frost98}. A few studies have been made to 
study $^{12}$C/$^{13}$C in molecular envelope of PNs
\citep[e.g.][]{palla00,balser02} and lend support to the existence of CBP.

Three $^{13}$C-bearing molecular species, $^{13}$CO, H$^{13}$CN,
and H$^{13}$CO$^+$, have been detected in our observations, allowing
us to estimate the $^{12}$C/$^{13}$C ratio. The 
integrated intensity ratio of the CO(2--1) and  $^{13}$CO(2--1)
is 25, yielding an abundance ratio of $N(^{12}{\rm CO})/N(^{13}{\rm CO})=73$.
Varying values of the   $^{12}$CO/$^{13}$CO ratio in \object{NGC\,7027} have been reported in the literature.  \citet{knapp85} obtained $40\la N(^{12}{\rm CO})/N(^{13}{\rm CO})\la130$.  \citet{sopka89}
did not detect the $^{13}$CO emission in \object{NGC\,7027} and estimated
$^{12}$CO/$^{13}$CO to be larger than 100.
\citet{kahane92}, \citet{bachiller97}, and \citet{josselin03} got a lower 
limit of 65, 25, and 11, respectively.  More recently, \citet{balser02} obtained
that
$N(^{12}{\rm CO})/N(^{13}{\rm CO})\sim31$, about half of our determination.

The HCN/H$^{13}$CN abundance ratio is 19, similar to the
$I$(HCN(3--2))/$I$(H$^{13}$CN(3--2)) integrated intensity ratio of 15.
{ The value may bear a larger error due to the weakness of 
the H$^{13}$CN(3--2) transition.}
This is for the first time that the isotopomer ratio of this species
in \object{NGC\,7027} is obtained.
The value is a factor of $\sim4$ lower than the 
$^{12}$CO/$^{13}$CO ratio. The higher $^{12}$CO/$^{13}$CO ratio
suggests that $^{13}$CO might be partly affected by 
selective photodissociation of the UV radiation field, as proposed
by \citet{knapp85}.  Since both the CO and HCN lines are likely to be optically thick, the above values may represent lower limits.   If the $^{12}$C/$^{13}$C ratio is indeed as low as suggested above, this may imply a non-standard mixing process such as CBP.

Alternatively,  the  HCO$^+$/H$^{13}$CO$^+$ ratio can be used.  The HCO$^+$ lines are weaker but are definitely optically thin.  We used the HCO$^+$(1--0) line at { 89.188\,GHz} and the H$^{13}$CO$^+$ (3--2) line at 260.255\,GHz
since the H$^{13}$CO$^+$(1--0) transition at 88.632\,GHz 
falsl within a spectral region which has a poor signal-to-noise ratio
and the HCO$^+$(3--2) emission at
267.558\,GHz falls out of our frequency ranges.
{ Our calculations yield a rather high HCO$^+$/H$^{13}$CO$^+$ abundance
ratio of 117 (see Table~\ref{col_n7027}). 
If the HCO$^+$ (1--0) transition is optical thick, the value is even higher.}
On the other hand, based on the $3\sigma$ upper
limit of the integrated intensity of the 
H$^{13}$CO$^+$(1--0) line, we estimate that 
$N({\rm HCO^+})/N({\rm H^{13}CO^+})>60$.
The HCO$^+$/H$^{13}$CO$^+$ ratio derived in the present study
is higher than the value of 40 given by \citet{hasegawa01}.
The abnormal carbon isotopic ratio derived from
HCO$^+$ probably suggests that chemical fractionation is
significant for this species.

\subsubsection{Oxygen}

During the AGB stage, the nucleosynthesis and dredge-up processes
cause destruction of $^{18}$O and enhancement of $^{17}$O at the
the surface of evolved stars \citep[see][for a recent review]{busso06}.
\citet{knapp85} detected the oxygen isotopic ratio for a sample of 
carbon-rich circumstellar envelopes and found 
the range of 300--800 and 300--1300 for $^{16}$O/$^{17}$O and $^{16}$O/$^{18}$O, respectively. 
Based on the fact that the $^{17}$O/$^{18}$O ratios in carbon-rich
envelopes are higher than that in the interstellar medium by
a factor of 4--5,
they ruled out AGB stars as the dominant source of oxygen in the 
interstellar medium.

From the detection of CO, C$^{17}$O, and C$^{18}$O emission in
\object{NGC\,7027}, we obtain the oxygen isotopic ratios
$N(^{16}{\rm O})/N(^{17}{\rm O})=698$, 
$N(^{16}{\rm O})/N(^{18}{\rm O})=1763$, 
and $N(^{17}{\rm O})/N(^{18}{\rm O})=3$. 
Our results are consistent with the ratios (or the lower limits) derived by 
\citet{knapp85} in \object{NGC\,7027}.
Compared with the oxygen isotopic ratios in the Sun \citep[
$N(^{16}{\rm O})/N(^{17}{\rm O})=0.2$,
$N(^{16}{\rm O})/N(^{18}{\rm O})=499$, and 
$N(^{17}{\rm O})/N(^{18}{\rm O})=2682$,][]{lodders03}, \object{NGC\,7027} clearly shows enhancement of
$^{17}$O and destruction of $^{18}$O. which well agrees with the
predictions of stellar models.

%HC$^{18}$O$^+$ has been detected in \object{NGC\,7027}. We obtain
%the integrated intensity ratio 
%$I$(HCO$^+$(1--0))/$I$(HC$^{18}$O$^+$(1--0))$=18$, yielding
%a HC$^{16}$O$^+$/HC$^{18}$O$^+$ ratio of 12, lower than
%the C$^{16}$O/C$^{18}$O ratio by a factor of two magnitude.
%The abnormal carbon and oxygen isotopic ratios derived from
%HCO$^+$ probably suggests that chemical fractionation is
%significant for this species.

\subsection{Hydrogen and helium recombination lines}

A total of over 30 H and He recombination lines are detected in our observations.  These include nine H$n\alpha$ lines, nine H$n\beta$ lines, and probably six H$n\gamma$ lines and eight He$n\alpha$ lines. The linewidths and integrated intensities of these lines are listed in Table~\ref{re_n7027}.

Due to optical depth effects, the line widths and radial
velocities of recombination lines at different frequencies may
be different \citep{ershov89}. The high frequency emission has negligible
optical depth, and is likely to be emitted from the whole ionized region. 
The low frequency emission, however, is more likely to be optically thick and 
if so will be emitted from the outer regions. Our data do not show any clear
correlation between the line frequencies and the line widths and
velocities, suggesting that these recombination lines are optically
thin. From profile fittings, we determine the averaged line widths ($FWHM$) for H$n\alpha$,
H$n\beta$ and He$n\alpha$ lines to be $44\pm8$\,km\,s$^{-1}$,
$54\pm9$\,km\,s$^{-1}$, and  $38\pm9$\,km\,s$^{-1}$, respectively.
No systematic difference is found for the widths of hydrogen lines
and the helium lines. The H$n\alpha$ lines detected here are
a factor of about 1.6
narrower than the low frequency H110$\alpha$ 
line detected by \citet{chaisson76}, which is expected to be optical thick.
The observed widths of hydrogen radio recombination lines can be interpreted 
by the model of \citet{ershov89} (cf. their Fig.2c).

The relative intensities of observed H$n\alpha$ lines as a function of their quantum numbers are plotted in Fig.~\ref{recombination}.
In this plot, we have corrected for the effect of beam dilution and translated the intensity unit into Jy. The low frequency lines
H76$\alpha$ and H110$\alpha$ reported by \citet{chaisson76}
are also plotted in this figure.
We have normalized the intensities of the H76$\alpha$ and H110$\alpha$
lines to $I({\rm H39\alpha})=1$ by assuming that they have the same 
beam-filling factor, which may introduce an error less than a factor of 5.
For comparison, we plot the theoretical predictions of
\citet{storey95} under the assumption of Case B with several combinations of 
electron temperatures and densities.  
%Fig.~\ref{recombination} shows that the relative intensities of the recombination lines are insensitive to the electron temperature and density.
Note that \citet{storey95} only gives the line emissivities for $n\le50$ and the predictions for $n>50$ in Fig.~\ref{recombination} are linear extrapolations of their results. 
If a correction for continuum free-free opacity is applied, the predicted
lines would have a larger slope than those shown in Fig.~\ref{recombination}. 
An inspection of Fig.~\ref{recombination} shows
that the intensities of the high-frequency H$n\alpha$ lines
($>50$\,GHz) are consistent with the theory of recombination and
are not affected by free-free opacity. The fact that the high frequency H$n\alpha$ lines are weaker than the theoretical predictions may suggest  that they are optically thick.

H$n\alpha$ and H$n\beta$ radio recombination lines were also
detected in molecular clouds \citep[e.g.][]{turner91,nummelin00},
allowing us to make a comparison study. The averaged main-beam temperatures for H$n\alpha$ lines
and H$n\beta$ lines determined from our observations are $T(\alpha)=0.046$\,K and $T(\beta)=0.014$\,K, respectively.  The corresponding $T_R(\alpha)/T_R(\beta)$ ratio in
\object{NGC\,7027} is 3.3, which is in good agreement with that found
in two molecular clouds \object{OMC-1} and \object{Sgr\,B2}
\citep{turner91}.  According to \citet{turner91}, stimulated emission must be significant
in order to explain such a $T_R(\alpha)/T_R(\beta)$ ratio. The observed $T_R(\alpha)$ value
in \object{NGC\,7027} is lower than that in \object{OMC-1} by a
factor of ten, which can be attributed to the smaller size ($\sim0.1$ pc) of this PN.
Following the calculations of \citet{turner91}, \object{NGC\,7027}
has an emission measure $EM\sim6\times10^6$\,cm$^{-6}$\,pc. 

\citet{thum98} studied the radio recombination lines of hydrogen in the
emission-line star \object{MWC\,349} and found that the amplification
factor of the recombination line maser has a peak value near $n=19$.
Fig.~\ref{recombination} compares hydrogen recombination lines in
\object{MWC\,349} and  those in \object{NGC\,7027}.
We can clearly see
that for \object{MWC\,349}, the intensities of the H$n\alpha$ lines near $n=30$ 
are consistently higher than the theoretical predictions, which has been
attributed to the broadband maser phenomenon \citep{thum98}.
The trend is not found in \object{NGC\,7027}. 
From the $T(\alpha)/T(\beta)$ ratio, we find stimulated emission to be significant for the hydrogen radio recombination in \object{NGC\,7027} and infer that all the H$n\alpha$ lines observed
in the present work have a similar amplification factor. 
\object{NGC\,7027} has a density of $~10^4$\,cm$^{-3}$, about a factor of $10^4$ lower than that of \object{MWC\,349}. Consequently, according to the model of \citet{thum98},
the quantum number of the peak amplification  ($n_{max}$) in 
\object{NGC\,7027} is larger than that in \object{MWC\,349}.
Since $n_{max}=19\pm2$ in \object{MWC\,349}, a similar amplification factor for the H$n\alpha$ would imply that  $n_{max}$  is in the range between 34--45 in \object{NGC\,7027}.

The hydrogen and helium radio recombination lines can also be used to calculate the He$^+$/H$^+$ abundance ratio. \citet{palmer69} deduced the He/H abundance ratio from radio recombination 
for a sample of \ion{H}{2} regions and found that the results
are in good agreement with those derived from optical data.
%The recombination coefficients of these helium radio recombination lines are unavailable to date.
Following \citet{palmer69} and assuming that the ratio of the integrated intensity of the He$n\alpha$ line to that of  the H$n\alpha$ represents the He$^+$/H$^+$ abundance ratio,
we obtain $N$(He$^+$)/$N$(H$^+$)$=0.11\pm0.02$. An additional
$20\%$ error is given by uncertainties in the measurements.
The derived He$^+$/H$^+$ abundance ratio is about a factor of two higher than that determined from the
optical spectra \citep{zhang05}.  
A possible explanation is that the He$^{+}$ zone is more extended than
the H$^+$ zone so that the effect of beam dilution is more severe
for the \ion{H}{1} radio recombination lines. To account for the 
discrepancy between the abundance derived from the radio data and
that from the optical data, the size of the He$^{+}$ zone may be larger
than  the H$^+$ zone by a factor of about 1.3. This is possible if
the ionization spectrum is hard enough and more UV photons
have an energy close to the ionization potential of helium.
In that case, the He$^{2+}$ zone is expected to be large in the high excitation
PNs. However, the \ion{He}{2} radio recombination lines are below our
detection limit, and thus we have not attempted to calculate the
elemental abundance ratio of helium to hydrogen.

\subsection{Line profiles}

\begin{figure*}
\includegraphics[height=20cm]{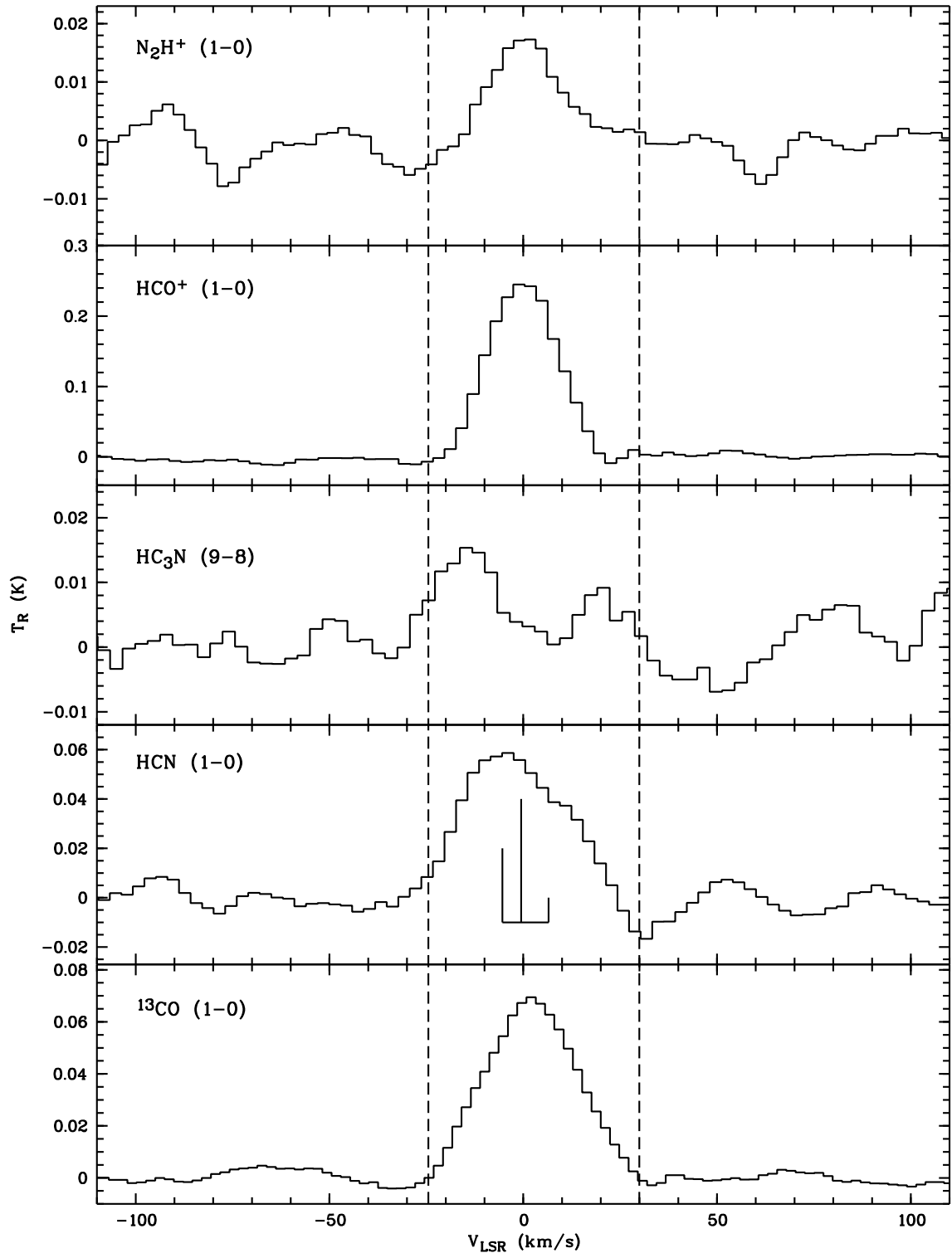}
\caption{Observed line profiles of NGC\,7027 by the ARO 12\,m telescope.
Positions and relative intensities of hyperfine components are marked
by vertical solid lines. { The spectral resolution is 1\,MHz.}
The velocity range of $^{13}$CO emission (above 1$\sigma$)
is indicated by vertical dashed lines.
}
\label{n7027_12m}
\end{figure*}

\begin{figure*}
\includegraphics[height=20cm]{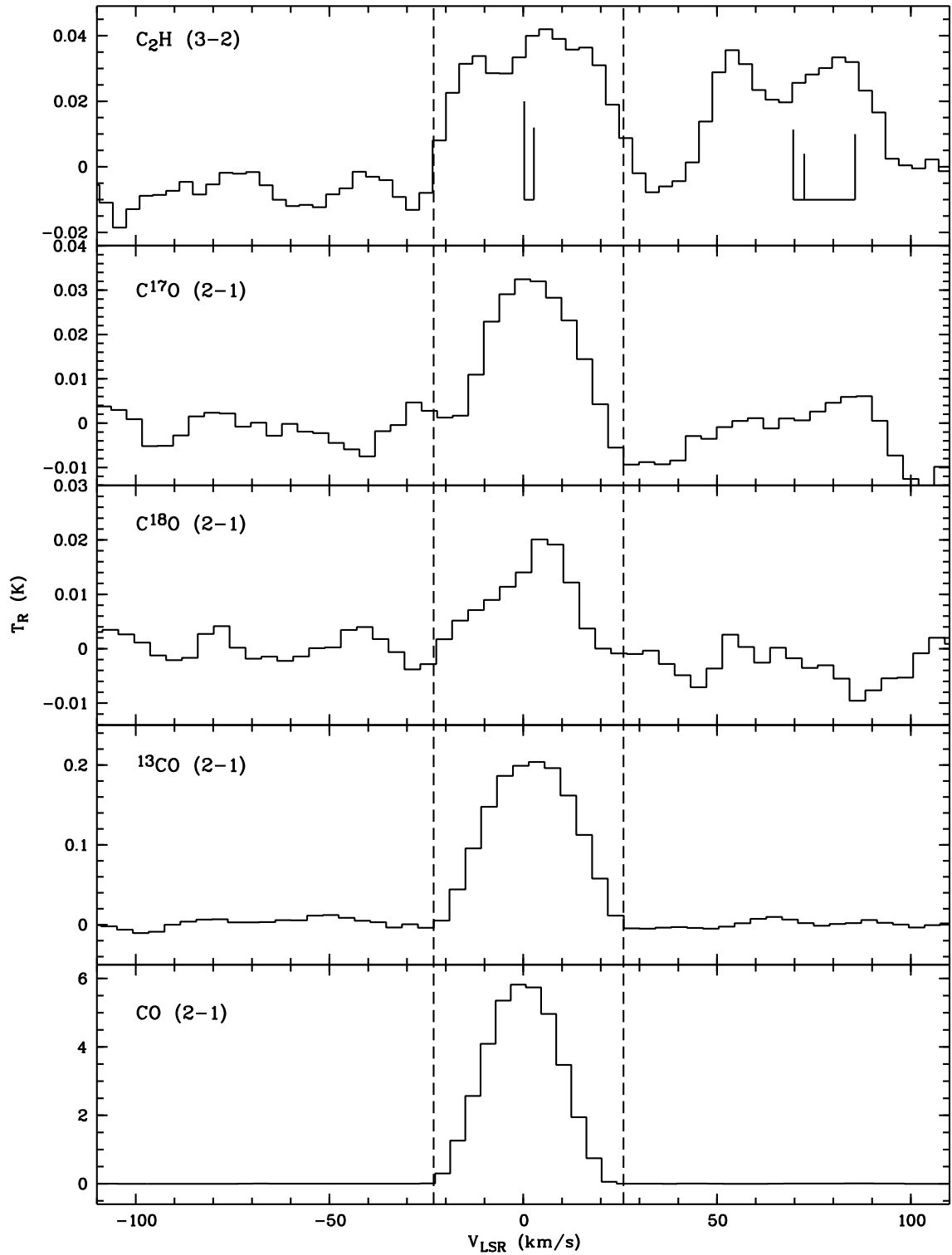}
\caption{Observed line profiles of NGC\,7027 by the SMT 10\,m telescope.
The description is same with those in Fig.~\ref{n7027_12m}
{ The spectral resolution is 3\,MHz.}\label{n7027_smt}}
\end{figure*}

\begin{figure*}
\epsscale{1}
\plotone{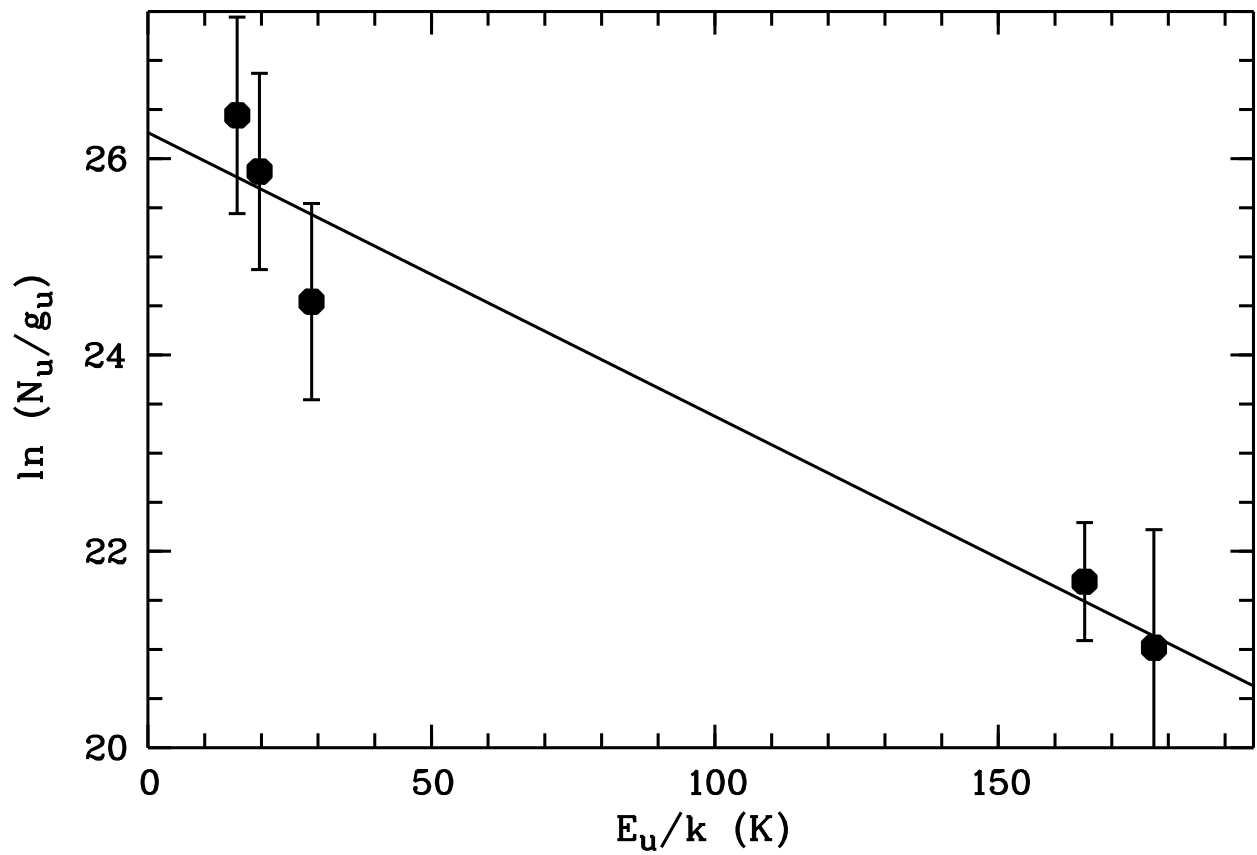}
\caption{Rotational diagram for HC$_3$N in NGC\,7027. The solid 
line is obtained through a linear least-squares fit of all the data.
}
\label{dia_n7027}
\end{figure*}

\begin{figure*}
\plotone{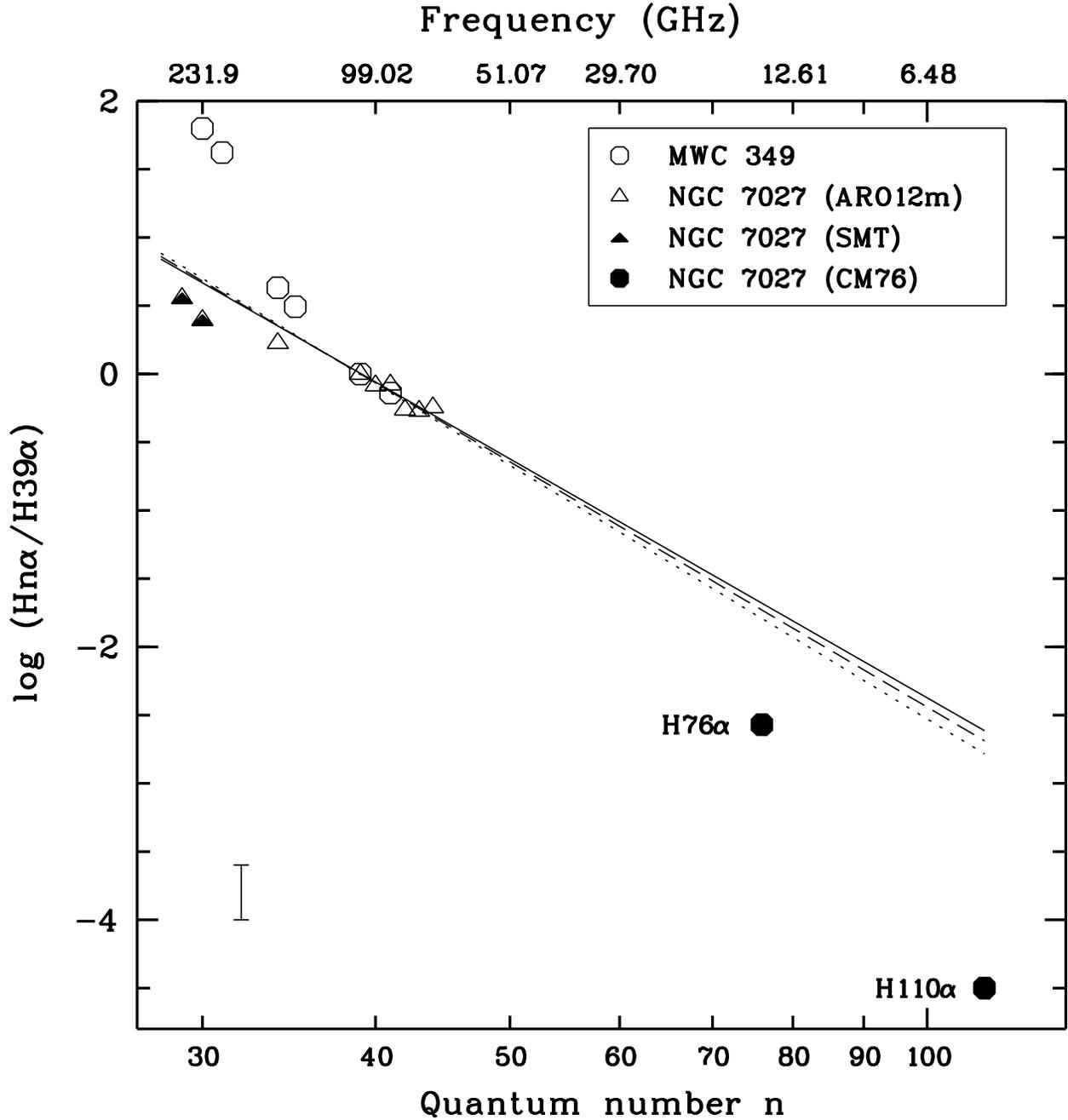}
\caption{Relative fluxes of \ion{H}{1} recombination $\alpha$-transitions
in MWC\,349 (open circles; Thum et al. 1998) and NGC\,7027. 
The open triangles and the filled triangles are from our observations
of  NGC\,7027
with the ARO\,12m and the SMT\,10m telescopes, respectively.
The error bar on the lower left indicates typical uncertainties of
our obervations.
The filled circles are from the observations of
NGC\,7027 by Chaison \& Malkan (1976).
The theoretical predictions of Storey and Hummer (1995) are given
by assuming $T_{\rm e}=12500$\,K and $N_{\rm e}=10^4$\,cm$^{-3}$ (solid
line),
$T_{\rm e}=12500$\,K and $N_{\rm e}=10^3$\,cm$^{-3}$ (dotted line), and
$T_{\rm e}=10000$\,K and $N_{\rm e}=10^4$\,cm$^{-3}$ (dashed line).
No correction for free-free continuum opacity is included.}
\label{recombination}
\end{figure*}

Fig.~\ref{n7027_12m} and Fig.~\ref{n7027_smt} present the line profiles
detected with the ARO\,12m telescope and the SMT\,10m telescope, respectively.
{ The HCO$^+$  line exhibits a self-absorption
feature at $V_{\rm LSR}\sim22$\,km/s, in accord with the observations of
HCO$^+$ line by \citet{deguchi90}. }
We also find a weak absorption for the HCN lines at $V_{\rm LSR}\sim30$\,km/s.
%No absorption feature is detected for the optically thin H$^{13}$CO$^+$ line.
%This is consistent with our finding that the column density of
%H$^{13}$CO$^+$ is much lower than those of HCO$^+$ and HC$^{18}$O$^+$.

{ Table.~\ref{mol_n7027} gives the $FWHM$ of detected molecular lines.
The emission lines from CO and its isotopomers have a width  between
23--30\,km/s.}
\citet{hasegawa01} found that the HCN, CN, and C$_2$H
have larger line widths than that of the CO emission, even
after taking into account the effects of hyperfine components.  Their
findings are confirmed by our observations. 
%Specially, we have also
%detected the HC$_3$N and C$_3$H$_2$ lines which show a similar
%trend. 
One possible explanation for the larger linewidths is that these molecular lines contain components arising from fast-moving clumps within the neutral envelope of \object{NGC\,7027}
\citep{redman03,huggins02}. The CO emission originates in a much more extended region
compared to other molecular species, and thus is less affected
by the contribution from the fast components. 
{ The present observations
show that the lines from HCO$^+$ and
N$_2$H$^+$ are consistently narrower than that from CO. 
We therefore infer that the fast clumps might be neutral and have no contribution to the emission from the molecular ions. The molecular ions mostly originate from the PDR, which has a lower expansion velocity. Further observations of lines from molecular ions
are required to verify this point.}
Fig.~\ref{n7027_12m} and Fig.~\ref{n7027_smt}
 show that the recombination lines are generally
broadener than CO lines, suggesting that the 
acceleration by the UV radiation field is significant 
in the ionized regions of \object{NGC\,7027}.

\subsection{Comparison with the spectra of IRC+10216 and CRL\,618}

{ 
Comparison with the spectra of AGB stars, PPNs, and PNs
can provide essential information of chemical evolution in late-type
stars.  Recent molecular line surveys of \object{IRC+10216}
and \object{CRL\,618} were presented by \citet{cernicharo00} and
\citet{pardo07}, respectively.  The frequency ranges are
129--172.5\,GHz for \object{IRC+10216},
and 80.25--115.75\,GHz, 131.25--179.25\,GHz, and 204.25--275.25\,GHz
for \object{CRL\,618}. Since these spectra were obtained using different 
instruments, it is hard to make a quantitative comparison with our observation
of \object{NGC\,7027}.  Through a qualitative comparison, we summarize
the main differences between the spectra of the three objects 
as follows:

\begin{list}{}{}
 \item[1.] The molecular lines detected in \object{NGC\,7027} are
much fewer and generally fainter than those in \object{IRC+10216} and
\object{CRL\,618}, even after the correction of the relative distances of the three objects.
This can be attributed to a) the low density in the neutral envelope of \object{NGC\,7027}, which makes it harder to
collisionally excite the higher rotational levels of
most molecules \citep{thronson86}, and b) destruction of molecules caused by the strong UV radiation and shock waves
in \object{NGC\,7027}.

\item[2.] Refractory Metal-bearing and silicon-bearing species which are
plentifully present in  \object{IRC+10216} have only week emission in
\object{CRL\,618} and are not detected in  \object{NGC\,7027}. These molecules may be depleted
onto dust grains with stellar evolution.

\item[3.] The CS emission is strong in
\object{IRC+10216}, medium in \object{CRL\,618}, and not seen in \object{NGC\,7027}. This might suggest
that shock waves play an important role in the chemistry of \object{NGC\,7027}.

\item[4.] Ionized species (HCO$^+$, HCS$^+$, and N$_2$H$^+$) and recombination lines detected in 
\object{CRL\,618} and \object{NGC\,7027} are not seen in the spectra of \object{IRC+10216}.
This is due to photoionization by the hotter central stars of \object{CRL\,618} and \object{NGC\,7027}.

\item[5.] The N$_2$H$^+$/HCO$^+$ abundance ratio in \object{NGC\,7027} is about a factor of 30 higher than
that in \object{CRL\,618}, supporting the hypothesis that H$_3^+$
is dominantly produced by soft X-ray from the central star.
\end{list}

The comparison between the spectra of \object{CRL\,618} and \object{NGC\,7027} is particularly 
meaningful because the former is a much evolved PPN and will evolve to become a young PN like
\object{NGC\,7027}
after only a few hundred years. From the above comparison, we can conclude that
the chemical compositions can dramatically change after the PPN stage.
}

\section{Summary}

In this spectral-line survey of the PN \object{NGC\,7027}, we have detected a
total of 67 spectral lines consisting of 21 molecular lines,
32 hydrogen and helium recombination lines, and 14 unidentified line.
%In particular, we detected HC$_3$N in a PN for the first time.
The line intensities were used to calculate excitation temperatures, 
column densities, and fractional abundances of the observed molecular
species. The main findings are summarized as follows:

\begin{list}{}{}
%\item[1.]  The rotation diagram of HC$_3$N  indicates that
%it may originates in two regions with excitation
%temperatures of 7\,K and 33\,K, respectively.

\item[1.] The low CN/HCN ratio seems to go against the theoretical expected photodissociation of HCN into CN in a strong UV environment.
This could imply that the photodissociation process might be compensated by the unexpectedly efficient reaction between CN and H$_2$.  Alternatively,  a large amount of HNC may have been transferred
into HCN in the hot envelope of \object{NGC\,7027}.

%\item[3.] The detection of HC$_3$N and C$_3$H$_2$ implies to the 
%presence of cold dense clumps, which shield the strong UV radiation from
%the central star.

\item[2.]  From the HCO$^+$/N$_2$H$^+$ abundance ratio,
we suggest that
the reaction of CO$^+$ and H$_2$ may be important for the production 
of HCO$^+$. The soft X-ray emission from the central star play
to important role for the production of 
N$_2$H$^+$ in \object{NGC\,7027}.

\item[3.]  We find evidence for underabundance of CS, suggesting that
the species may have been destroyed by shocks in the young PN.

\item[4.] Although the estimated CO/$^{13}$CO, HCN/H$^{13}$CN,
and HCO$^+$/$^{13}$CO$^+$ ratios are different, the resulting
low $^{12}$C/$^{13}$C ratio
supports the presence of extra mixing process in AGB stage.
The oxygen isotopomer ratios is consistent with the predictions 
of stellar models.

\item[5.] The intensities of hydrogen recombination lines suggest
that stimulated excitation may be significant. Our estimates of
He$^+$/H$^+$ abundance ratio from the ratio recombination lines
is higher than that from the optical lines. The discrepancy can
be explained if the He$^+$ zone is more extended that
the H$^+$ zone in the young PN.

\item[6.] The different widths of lines from CO, other neutral molecules, molecular ions, and hydrogen and helium can be used as dynamical probes of the circumstellar envelope of \object{NGC\,7027}.
Our results suggest the presence of fast neutral clumps within
the molecular envelope of \object{NGC\,7027}.

\end{list}

These observations show that the circumstellar envelope of \object{NGC\,7027} can serve as a very useful chemical laboratory.  Comparison with existing theoretical chemical models suggests that a more comprehensive model is required to explain the observed molecular abundances, in particular for
the production of N$_2$H$^+$,  the destruction of CS, and the survival of
HC$_3$N and C$_3$H$_2$ in the PN.   A more comprehensive model will need to include 
effects of soft X-ray emission from the central star, shocks, and dense clumps.

Changing physical conditions are the main factor for the different molecular abundances in the
circumstellar envelopes of stars at different evolutionary stages.
The observational results reported in the present paper provide a foundation for further studies of circumstellar chemistry.
In this spectral survey program, we also obtained radio spectra of two AGB stars, \object{IRC+10216} and \object{CIT\,6}, and one proto-PN, \object{CRL\,2688}, in addition to the PN \object{NGC\,7027}.
A systematic comparison of the chemical compositions in different evolutionary stages will be presented in a forthcoming paper.

\acknowledgments

We  thank Aldo Apponi for his help in the processing of the ARO data. 
We also thank Jun-ichi Nakashima and Jin-Hua He for useful discussions.
This work is supported in part by a grant awarded to SK from the Research Grants Council of Hong Kong.  DVT acknowledges support of this work by Academia Sinica, Taiwan, and the Natural Science Council of Taiwan.

\clearpage

\begin{deluxetable}{llrcllll}
\tabletypesize{\scriptsize} 
\tablecaption{Detected molecular transitions.
\label{mol_n7027}}
\tablewidth{0pt}
\tablehead{
\colhead{Species} & \colhead{Transition} & \colhead{Frequency}& \colhead{$rms$}& \colhead{$T_{\rm R}^{~a}$} 
&\colhead{$\int T_{\rm R}$d$v^{~a}$} &\colhead{$FWHM^{~a}$}   
& \colhead{Remarks$^b$}\\
 & \colhead{(upper--lower)} & \colhead{(GHz)} & \colhead{(mK)} & \colhead{(K)} & \colhead{(K~km/s)} & \colhead{(km/s)}\\
}
\startdata
CO          &     2--1               & 230.538 & 3.0& 5.824 &    141.97 & 24.2 &     H01,T86,B97  \\
$^{13}$CO   &     1--0               & 110.201 & 3.3& 0.069 &      1.95 & 27.5 &    S89,B97  \\
            &     2--1               & 220.398 & 3.7& 0.204 &      5.72 & 29.1 &   H01,J03,B97   \\
C$^{18}$O   &     2--1               & 219.560 & 3.6& 0.020 &      0.41 & 23.3 &    \\
C$^{17}$O   &     2--1               & 224.714 & 4.3& 0.032 &      0.86 & 25.4 &        \\
CN          &  2--1,3/2--3/2         & 226.314 & 3.9& 0.028 &      1.79 & \nodata&   *,C93,J03,B97   \\
            &  2--1,3/2--1/2         & 226.659 & 3.9& 0.091 &      4.95 & 44.5 &   *,C93,J03,B97   \\
            &  2--1,5/2--3/2         & 226.874 & 3.9& 0.197 &      7.76 & 36.5 &    *,C93,J03,B97   \\
            &  2--1,5/2--1/2         & 227.192 & 3.3& 0.008:  &      0.26:$^c$& \nodata &      \\
C$_2$H      &  3--2,7/2--5/2         & 262.004 & 5.5& 0.041   &      2.01 & 94.1 &   *,H01  \\
            &  3--2,5/2--3/2         & 262.067 & 5.5& 0.033   &      1.74 & 96.7 &    *,H01 \\
C$_3$H$_2$  &  4$_{32}$--3$_{21}$    & 227.169 & 3.3& 0.008:  &  0.26:$^c$ & \nodata &   \\
            &  6$_{25}$--5$_{14}$    & 251.527 & 4.0& 0.010:  &    0.19:   & \nodata &         \\
            &  5$_{33}$--4$_{22}$    & 254.988 & 4.1& 0.012:  &    0.20:  &  \nodata &        \\
HCN         &     1--0               & 88.632  & 3.8& 0.059   &      1.94 &  36.6    & D86,D90,S89,C93,J03,B97\\
            &     3--2               & 265.886 & 5.5& 0.207   &      8.22 &  56.6    & H01,C93   \\
H$^{13}$CN  &     1--0               & 86.339  & 3.5& 0.009:  &      0.28:&  \nodata & \\ 
            &     3--2               & 259.012 & 4.3& 0.013   &      0.56&   64.9:  &  \\
HCO$^+$     &     1--0               & 89.188  & 3.7& 0.247   &      5.39 &  19.2   &  S89,D90,C93,J03,B97 \\
H$^{13}$CO$^+$&   3--2               & 260.255 & 4.1& 0.017   &      0.27 &  16.1   &   H01   \\
HC$^{17}$O$^+$&   1--0               & 87.057  & 3.4& 0.008:  &      0.21:&  \nodata &   \\
HC$^{18}$O$^+$&   1--0               & 85.162  & 7.4& 0.015:  &      0.30:&  \nodata & \\
HCS$^+$     &     6--5               & 256.028 & 4.0& 0.014   &      0.25 &  17.0:   & H01   \\
HC$_3$N     &     8--7               & 72.783  & 3.6& 0.014   &      0.34 &  \nodata & \\
            &     9--8               & 81.881  & 3.6& 0.015   &      0.31 &  \nodata    &  \\
 
            &    10--9               & 90.979  & 3.1& 0.008:  &      0.16:&  \nodata &     \\
            &    11--10              & 100.076 & 2.9& 0.011   &      0.22 &  \nodata & \\
            &    12--11              & 109.174 & 2.8& 0.004:  &      0.12:& \nodata  &\\
            &    26--25              & 236.513 & 3.4& 0.004:  &      0.11:& \nodata         \\
            &    27--26              & 245.606 & 2.7& 0.014   &      0.23 &  \nodata &      \\
            &    28--27              & 254.700 & 4.1& 0.011:  &      0.13:&  \nodata &   \\
N$_2$H$^+$  &     1--0               &  93.173 & 2.6& 0.017   &      0.35 &  19.7    & C93,J03 \\
%            &                        &  105.092& 3.4& 0.012   &      0.25 &           \\
%            &                        &  105.791& 3.4& 0.013   &      0.34 &           \\
%            &                        & 221.463 & 3.8& 0.048   &      1.45 &           \\
%            &                        & 230.601 &  0.008   &      0.33 &   
\enddata
\tablenotetext{a}{the symbol ":" indicates uncertain detections.}
\tablenotetext{b}{Also detected by: B97--Bachiller et al. (1997);
C93--Cox et al. (1993); D86--Deguchi et al. (1986);
D90--Deguchi et al. (1990); J03--Josselin \& Bachiller (2003);
H01--Hasegawa \& Kowk (2001); S89--Sopka et al. (1989);
T86--Thronson \& Bally (1986). *--unsolved fine-structure lines.}
\tablenotetext{c}{The CN at 227.192\,GHz and the C$_3$H$_2$ at 227.169\,GHz lines
are blended with each other.}
\end{deluxetable}

\clearpage

\begin{deluxetable}{lccc}
%\tabletypesize{\scriptsize}
\tablecaption{Unidentified features.
\label{un_n7027}}
\tablewidth{0pt}
\tablehead{
\colhead{Frequency}& \colhead{$rms$}& \colhead{$T_{\rm R}$} &
\colhead{$\int T_{\rm R}$d$v$} \\
 \colhead{(GHz)} & \colhead{(mK)} & \colhead{(K)} & \colhead{(K~km/s)}  \\
}
\startdata
74.739 & 3.8& 0.012   &      0.63             \\
89.656$^a$ & 3.6& 0.012   &      0.26  \\
90.739$^a$ & 3.1& 0.013   &      0.23   \\
95.246 & 3.2& 0.011   &      0.47            \\
96.236$^a$& 5.3& 0.019   &      0.35  \\
98.079& 2.9& 0.010   &      0.20            \\
102.789& 3.3& 0.013   &      0.29            \\
107.802& 2.8& 0.012   &      0.37            \\
229.256$^b$ & 3.7& 0.015   &      0.22     \\
257.637$^b$ & 4.8& 0.015   &      0.25    \\
261.569$^b$ & 5.5& 0.020   &      0.28            \\
261.820 & 5.5& 0.019   &      0.36            \\
263.369$^b$ & 5.0& 0.017   &      0.26   \\
264.744 & 5.2& 0.019   &      0.36          \\
\enddata
\tablenotetext{a}{ Also detected in Sagittarius B2 (M) \citep{tur89};}
\tablenotetext{b}{ Also detected in Sagittarius B2 (N) \citep{num98}.}
\end{deluxetable}

\begin{deluxetable}{llrllcc}
\tabletypesize{\scriptsize}
\tablecaption{Non-detected molecular species in NGC\,7027.
\label{limit}}
\tablewidth{0pt}
\tablehead{
\colhead{Species} & \colhead{Transition} & \colhead{Frequency}& \colhead{$rms$} &
\multicolumn{2}{c}{$N$(cm$^{-2}$)}&  \colhead{$N$(obs.)/} \\
\cline{5-6}
 & \colhead{(upper--lower)} & \colhead{(GHz)} & \colhead{(mK)} &
 \colhead{obs.$^a$} & \colhead{model$^b$} &
\colhead{$N$(model)$^a$} \\
}
\startdata
CS      & $\nu$=0 J=5--4    & 244.936  & 2.7  & 1.9e12 & 2.5e13 & 0.08\\
C$_3$H  &  $^2{\Pi}_{3/2}$ J=23/2--21/2 b& 263.332  & 4.9  & 1.1e13&  ...   & ...  \\
C$_4$H  & N=24--23 a        & 228.349  & 2.9  & 1.7e13 & 1.4e11 & 121 \\
SiC     & $^3{\Pi}_2$ J=6--5& 236.288  & 3.5  & 7.5e12 & 2.6e12 & 2.9 \\
SiO     & $\nu$=0 J=6--5    & 260.518  & 5.2  & 1.0e13 & 1.5e13 & 0.7 \\
SiS     & $\nu$=0 J=14--13  & 254.103  & 3.7  & 8.9e13 &  ...   & ... \\
HNC     &  J=1--0           & 90.664   & 3.4  & 1.4e12 &  ...   & ... \\
HC$_5$N &  J=27--26         & 71.890  & 3.6  &  1.5e12&  ...   &  ... \\
CH$_3$CN&  J$_{\rm K}$=5(1)--4(1),5(0)--4(0)       & 91.986  & 2.4 & 2.3e13& ...&...\\
H$_2$CO &  3(1,2)-2(1,1)    & 225.678  &  3.9   & 4.5e12 & 9.2e9  & 489 \\ 
N$_2$D$^+$ & J=1--0           & 77.109   & 4.1  & 2.8e12 &  ...   & ... \\
\enddata
\tablenotetext{a}{Upper limits.}
\tablenotetext{b}{Calculated by $\int n(i)4\pi r^2dr/\Delta S$
from Hasegawa et al. (2000), where $\Delta S$ is the projected source area.}
\end{deluxetable}

\begin{deluxetable}{lrrll}
\tabletypesize{\scriptsize} 
\tablecaption{Recombination lines in NGC\,7027.
\label{re_n7027}}
\tablewidth{0pt}
\tablehead{
\colhead{Line} &  \colhead{Frequency}& \colhead{$T_{\rm R}$} &\colhead{$\int T_{\rm R}$d$v$} &\colhead{$FWHM$} \\
 & \colhead{(GHz)} & \colhead{(K)} & \colhead{(K~km/s)} & \colhead{(km/s)}\\
}
\startdata
H44$\alpha$ &   74.645  &  0.045 & 2.01 & 46.6\\
H43$\alpha$ &   79.913  &  0.037 & 1.77 & 52.5\\
H42$\alpha$ &   85.688  &  0.037 & 1.68 & 35.3\\
H41$\alpha$ &   92.034  &  0.054 & 2.39 & 41.2\\
H40$\alpha$ &   99.022  &  0.051 & 2.19 & 43.5\\
H39$\alpha$ &  106.737  &  0.049 & 2.47 & 49.4\\
H34$\alpha$ &  160.212  &  0.048 & 2.77 & 40.6\\
H30$\alpha$ &  231.901  &  0.037 & 1.88 & 46.2\\
H29$\alpha$ &  256.302  &  0.053 & 2.46 & 44.3\\
 &  &  &  & \\
H55$\beta$  &   74.940  &  0.013 & 0.46 & 52.9\\
H54$\beta$  &   79.104  &  0.025 & 0.68 & 35.7\\
H53$\beta$  &   83.582  &  0.011 & 0.41 & 48.5\\
H52$\beta$  &   88.406  &  0.016 & 0.60 & 60.1\\
H51$\beta$  &   93.607  &  0.010 & 0.40 & 64.5\\
H50$\beta$  &   99.225  &  0.013 & 0.53 & 49.8\\
H49$\beta$  &  105.302  &  0.012 & 0.57 & 65.5\\
H37$\beta$  &  240.021  &  0.013 & 0.54 & 57.5\\
H36$\beta$  &  260.033  &  0.011 & 0.42 & 92.1:\\
&&&&\\
H61$\gamma$ &   80.900  &  0.012 & 0.48 & 50.6\\
H60$\gamma$ &   84.914  &  0.024 & 0.29:& 17.8:\\
H58$\gamma$ &   93.776  &  0.009 & 0.28 & 64.1\\
H57$\gamma$ &   98.671  &  0.006 & 0.16:& 31.0:\\
H56$\gamma$ &  103.915  &  0.009 & 0.22:& 35.5:\\
H55$\gamma$ &  109.536  &  0.008 & 0.25 & 35.1\\
&&&&\\
He44$\alpha$&   74.674  &  0.009 & 0.21 & 40.2\\
He43$\alpha$&   79.945  &  0.012 & 0.17:& 33.3:\\
He42$\alpha$&   85.723  &  0.023 & 0.37:& 36.3:\\
He41$\alpha$&   92.072  &  0.012 & 0.28 & 27.0\\
He40$\alpha$&   99.063  &  0.008 & 0.22 & 55.5:\\
He39$\alpha$&  106.781  &  0.005 & 0.18: & 31.6:\\
He30$\alpha$&  231.995  &  0.009 & 0.23 & 42.0\\
He29$\alpha$&  256.406  &  0.014 & 0.35 & 35.2\\
\enddata
\end{deluxetable}

\begin{deluxetable}{lrcllll}
\tablecaption{Excitation temperatures ($T_{\rm ex})$, column densities ($N$), and
abundances relative to H$_2$ ($f_{\rm X})$.
\label{col_n7027}}
\tablewidth{0pt}
\tablehead{
\colhead{Species} & \colhead{$T_{\rm ex}$(K)} &
\multicolumn{2}{c}{$N$(cm$^{-2}$)}& \colhead{$N$(obs.)/} &  \colhead{${f_{\rm X}}^b$} & \colhead{N/$f_{\rm X}$(cm$^{-2}$)}\\
\cline{3-4} 
\colhead{} & \colhead{} & \colhead{obs.} & \colhead{model$^a$}
&  \colhead{$N$(model)}
}
\startdata
HC$_3$N       & 34.6  &  4.39e13 &...&...       & 8.5e-9 &5.2e21\\
CN            &       &  1.02e14 &2.4e14 &0.43  & 7.0e-8 &1.5e21\\
C$_2$H        &       & 3.49e14  &1.5e14 &2.33  & 5.4e-8 &6.5e21\\
C$_3$H$_2$    &       &  1.11e14:&...    &...   & 8.3e-9:&1.3e22:\\
HCS$^+$       &       &  4.96e12 &9.0e11 &5.51  & 1.0e-9 &5.0e21\\
N$_2$H$^+$    &       &  2.43e13 &3.2e5  &7.59e7& 3.8e-9 &6.4e21\\
HCO$^+$       &       &  3.38e14 &4.6e13 &7.35  & 4.8e-8 &7.0e21\\
H$^{13}$CO$^+$&       &  7.23e11 &...&...       & 4.1e-10&1.8e22\\
HC$^{17}$O$^+$&       &  2.85e13:&...&...       & 2.3e-9:&1.2e22:\\
HC$^{18}$O$^+$&       &  3.89e13:&...&...       & 3.2e-9:&1.2e22:\\
HCN           &  5.4  &  7.05e14 &1.8e14 &3.92  & 4.5e-8 &1.6e22\\
H$^{13}$CN    &       &  3.65e13 &...&...       & 3.9e-9 &9.4e21\\
CO            &  40.1 &  3.79e18 &3.2e20 &0.01  & 1.1e-4 &3.4e22\\
$^{13}$CO     &       &  5.19e16 &...&...       & 7.4e-6 &7.0e21\\
C$^{18}$O     &       &  2.15e15 &...&...       & 3.8e-7 &5.7e21\\
C$^{17}$O     &       &  5.43e15 &...&...       & 7.4e-7 &7.3e21\\
\enddata
\tablenotetext{a}{Calculated by $\int n(i)4\pi r^2dr/\Delta S$
from Hasegawa et al. (2000), where $\Delta S$ is the projected source area.}
\tablenotetext{b}{For the species with optically thick emission, 
this is a lower limit (see text for details).}
\end{deluxetable}

\clearpage

\clearpage

\clearpage

\clearpage
{\plotone{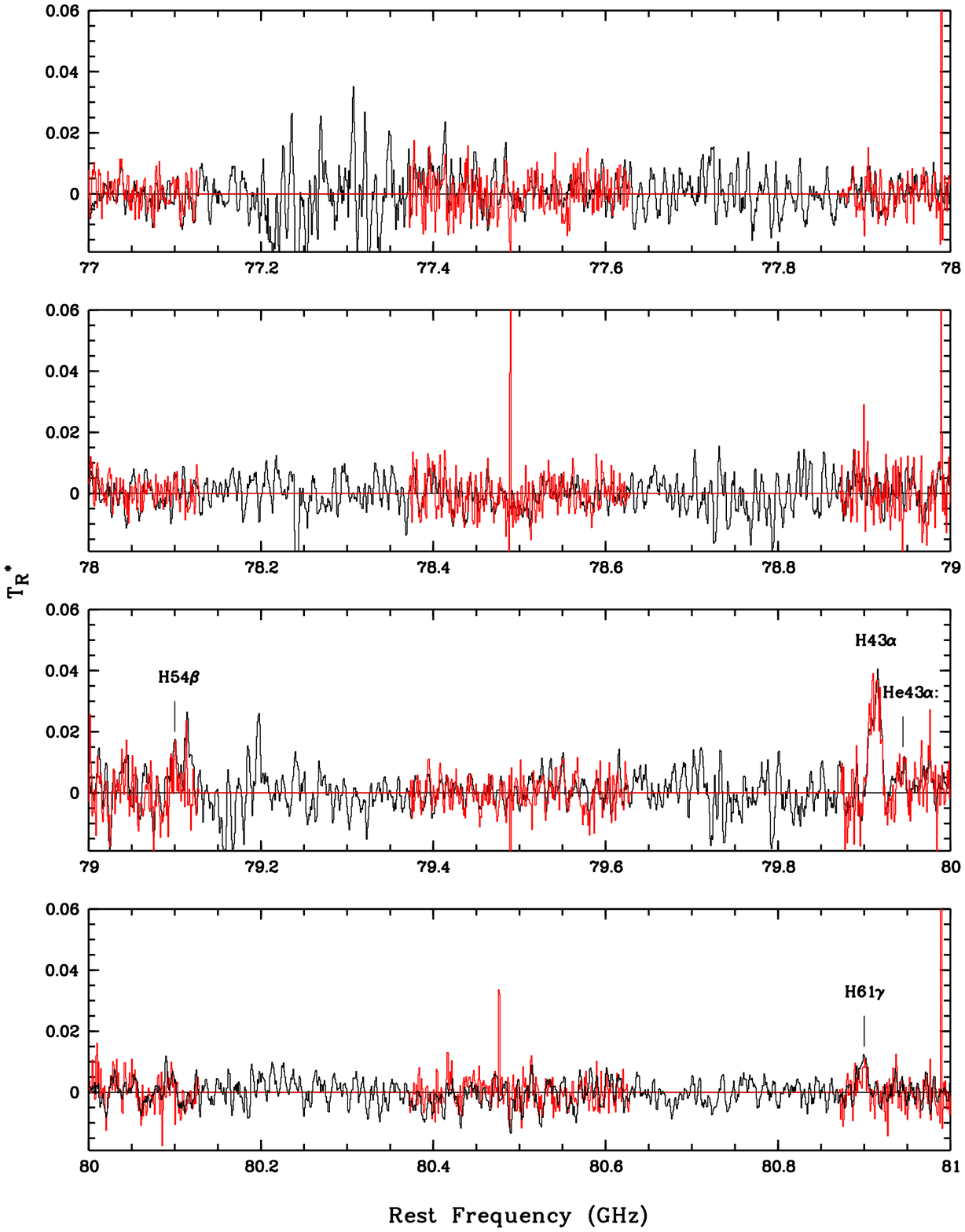}}\\
\centerline{Fig. 3. --- Continued.}
\clearpage
{\plotone{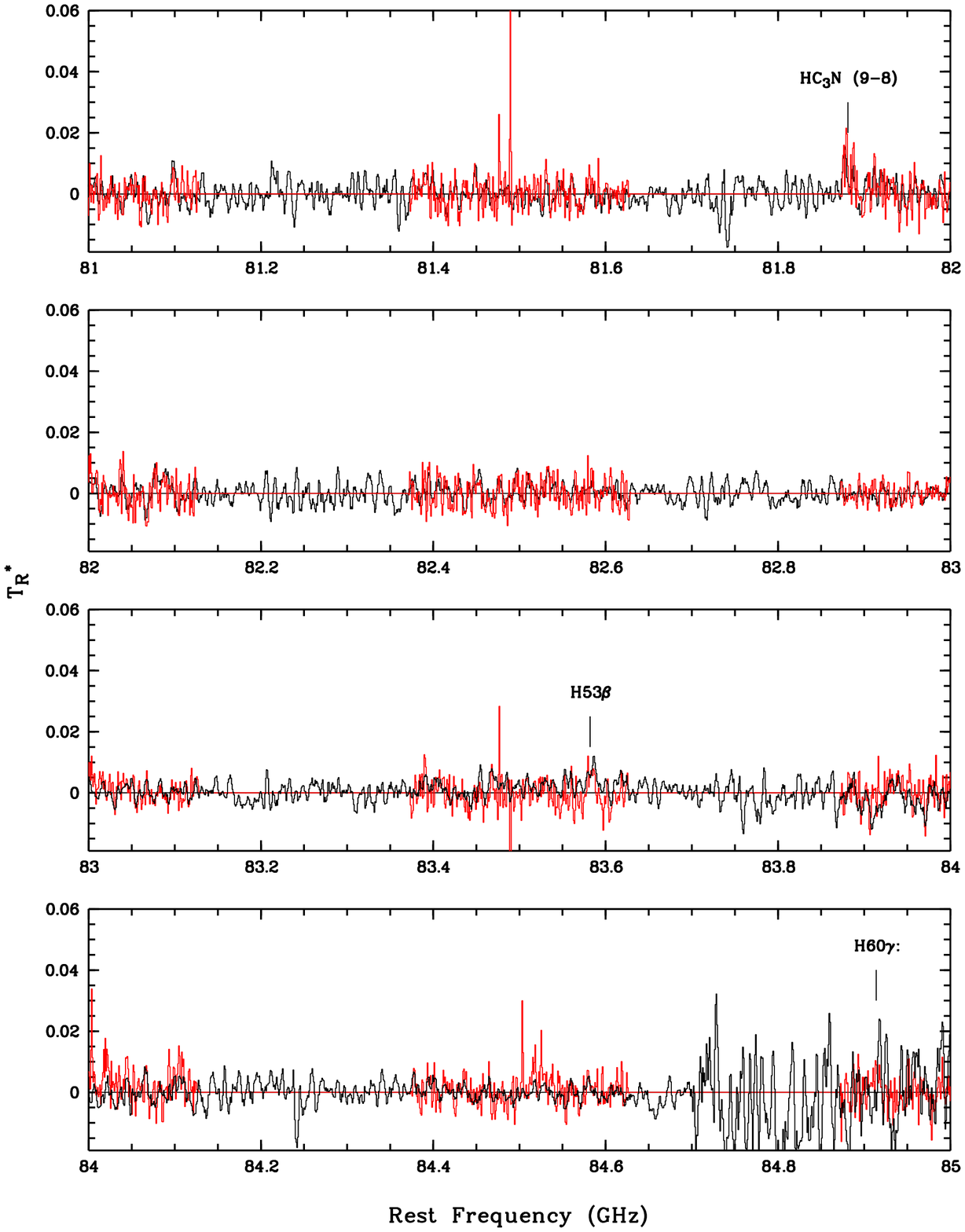}}\\
\centerline{Fig. 3. --- Continued.}
\clearpage
{\plotone{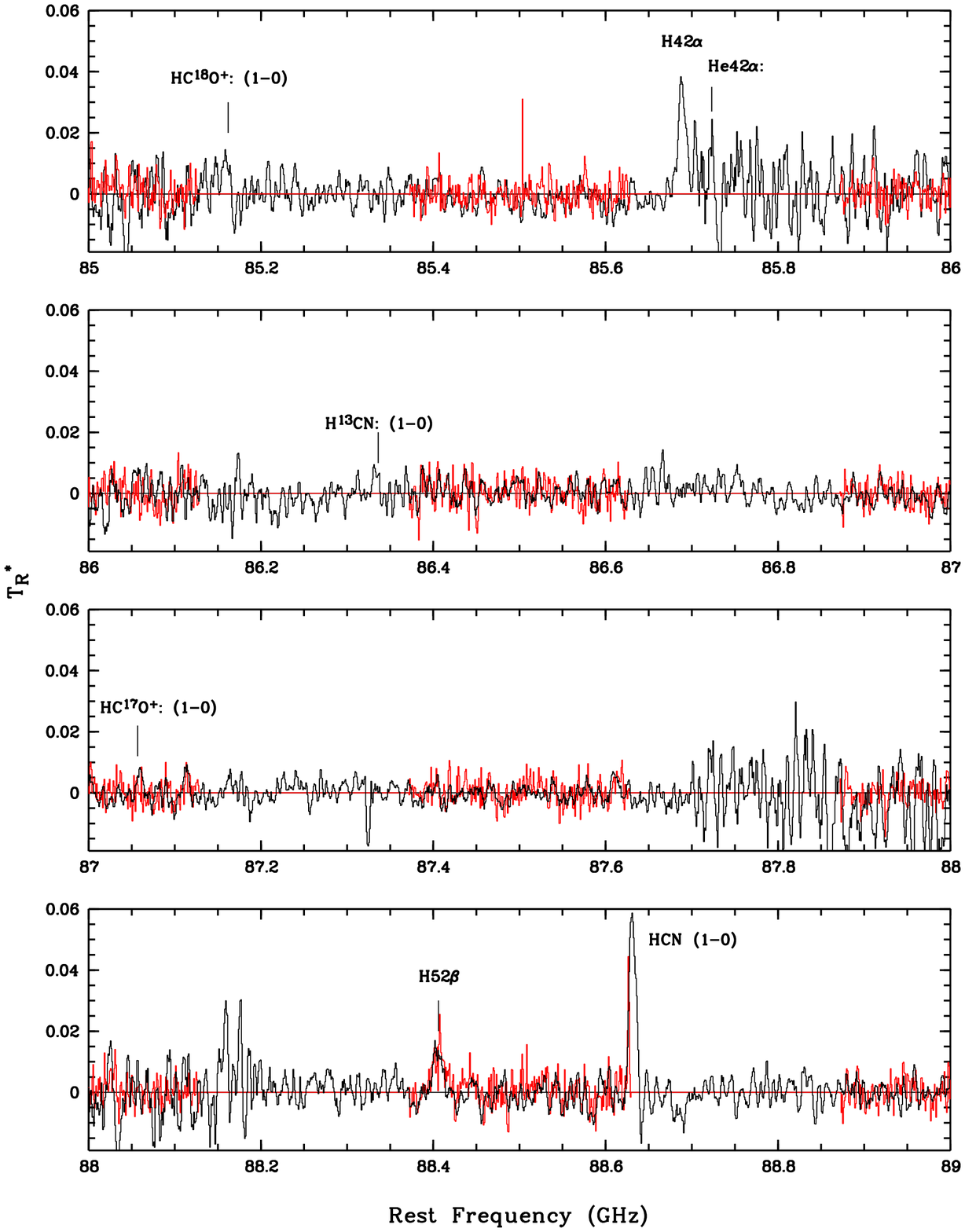}}\\
\centerline{Fig. 3. --- Continued.}
\clearpage
{\plotone{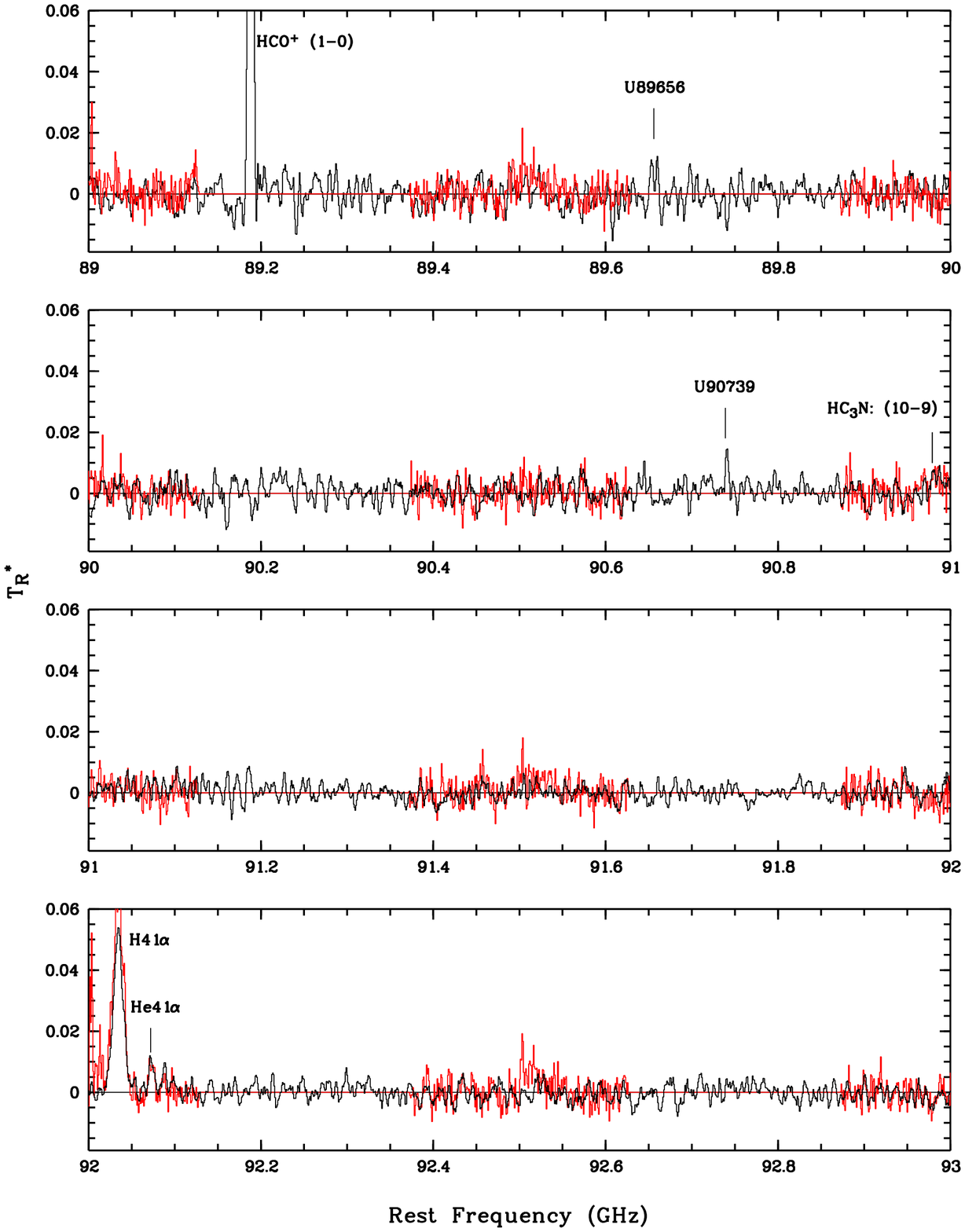}}\\
\centerline{Fig. 3. --- Continued.}
\clearpage
{\plotone{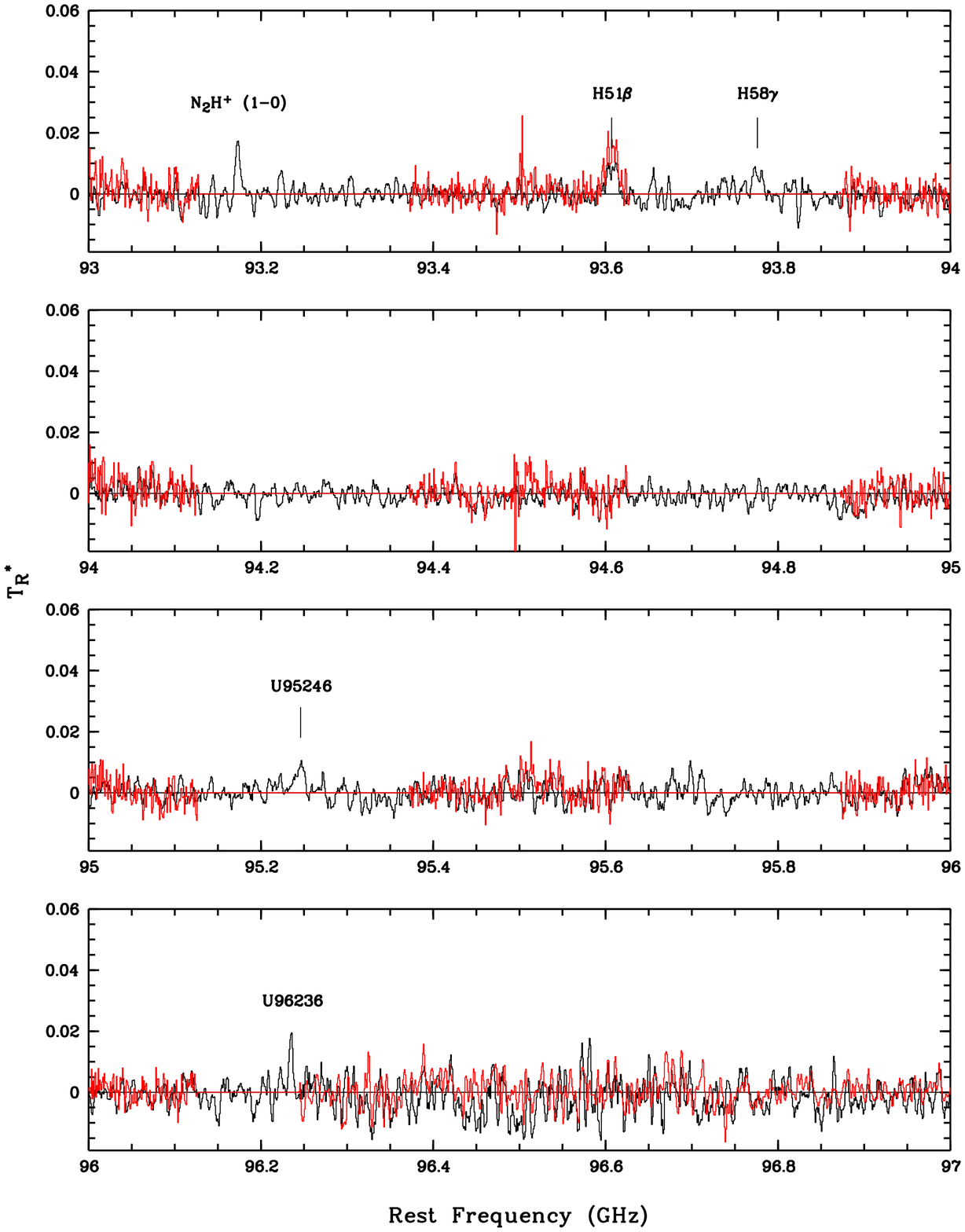}}\\
\centerline{Fig. 3. --- Continued.}
\clearpage
{\plotone{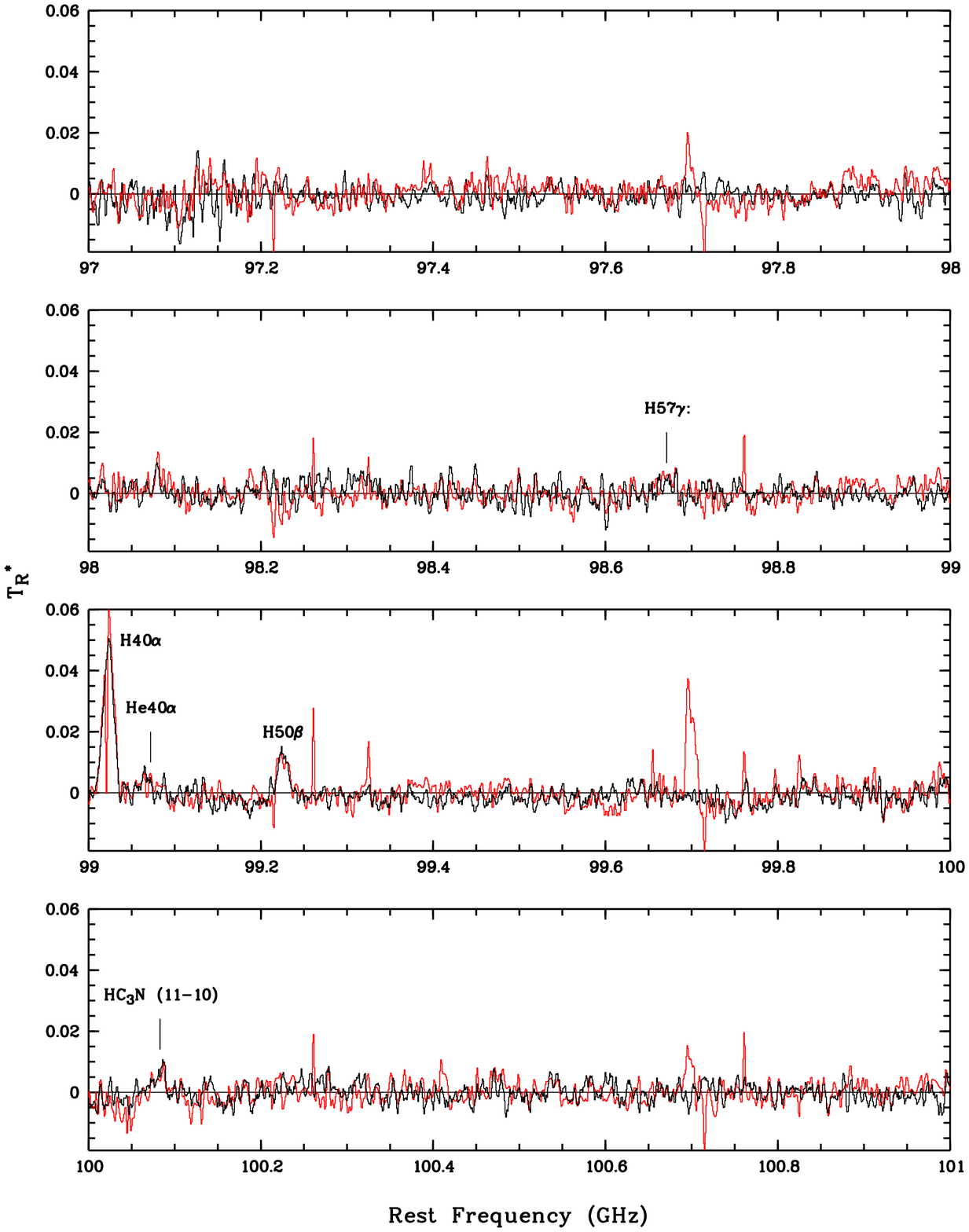}}\\
\centerline{Fig. 3. --- Continued.}
\clearpage
{\plotone{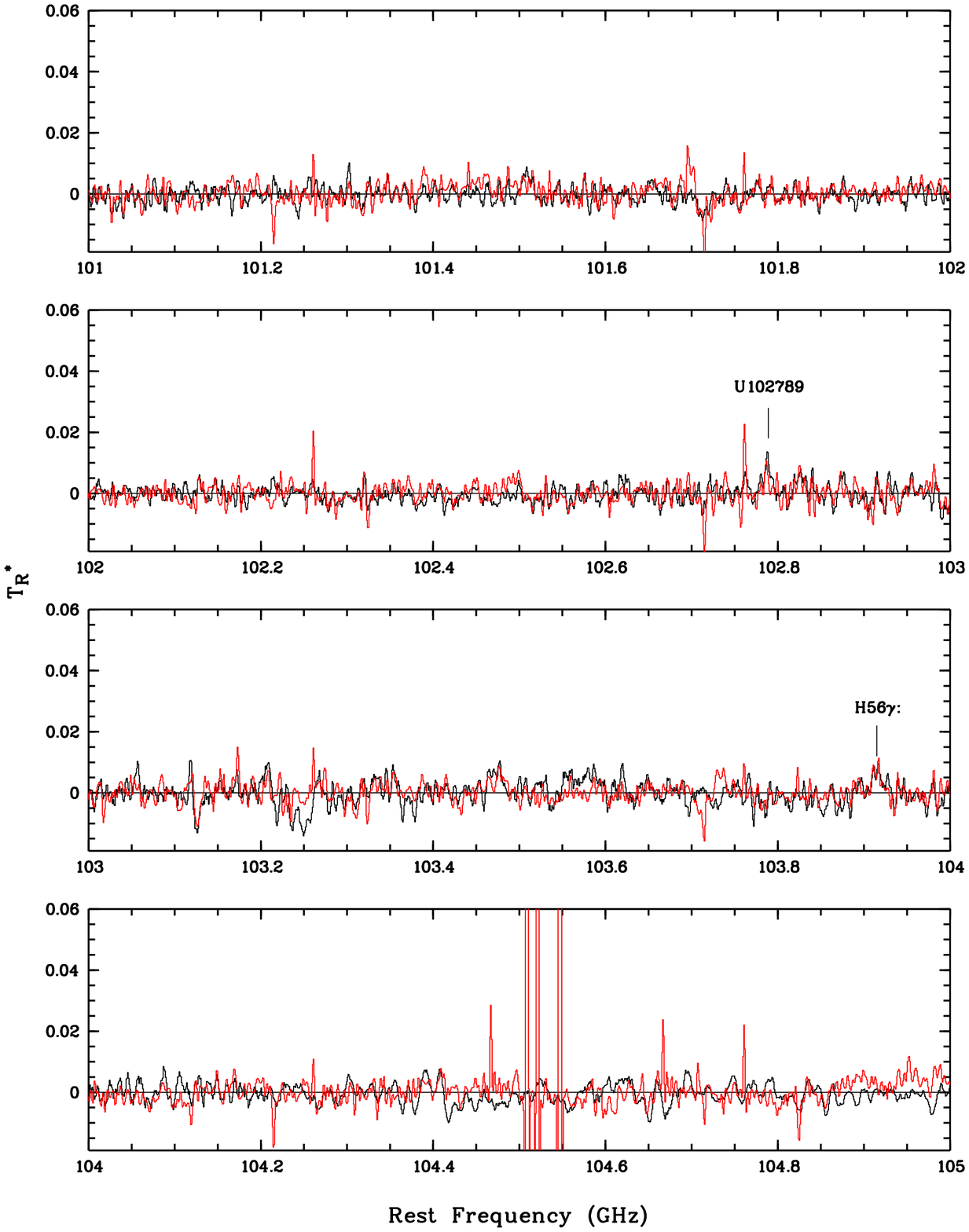}}\\
\centerline{Fig. 3. --- Continued.}
\clearpage
{\plotone{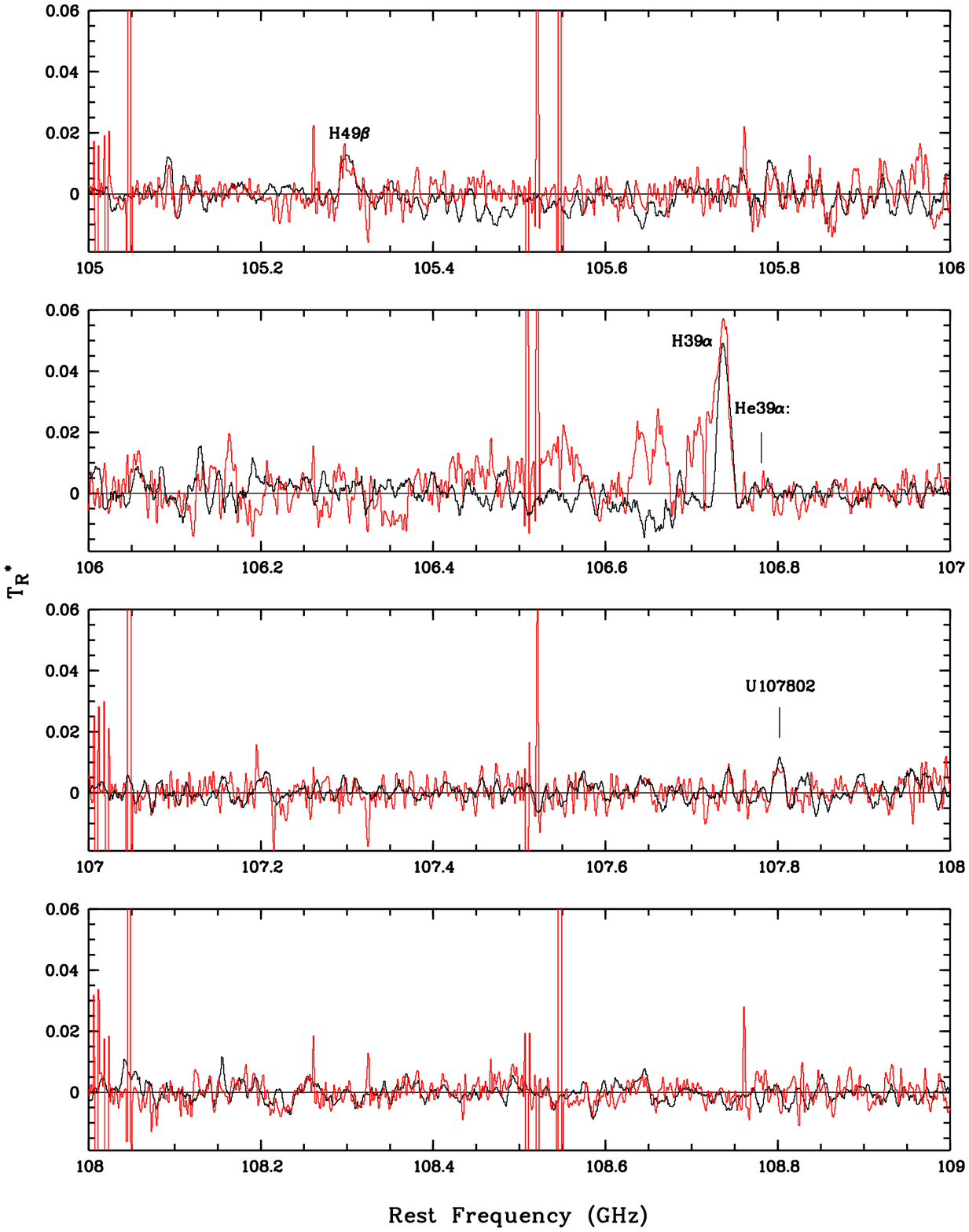}}\\
\centerline{Fig. 3. --- Continued.}
\clearpage
{\plotone{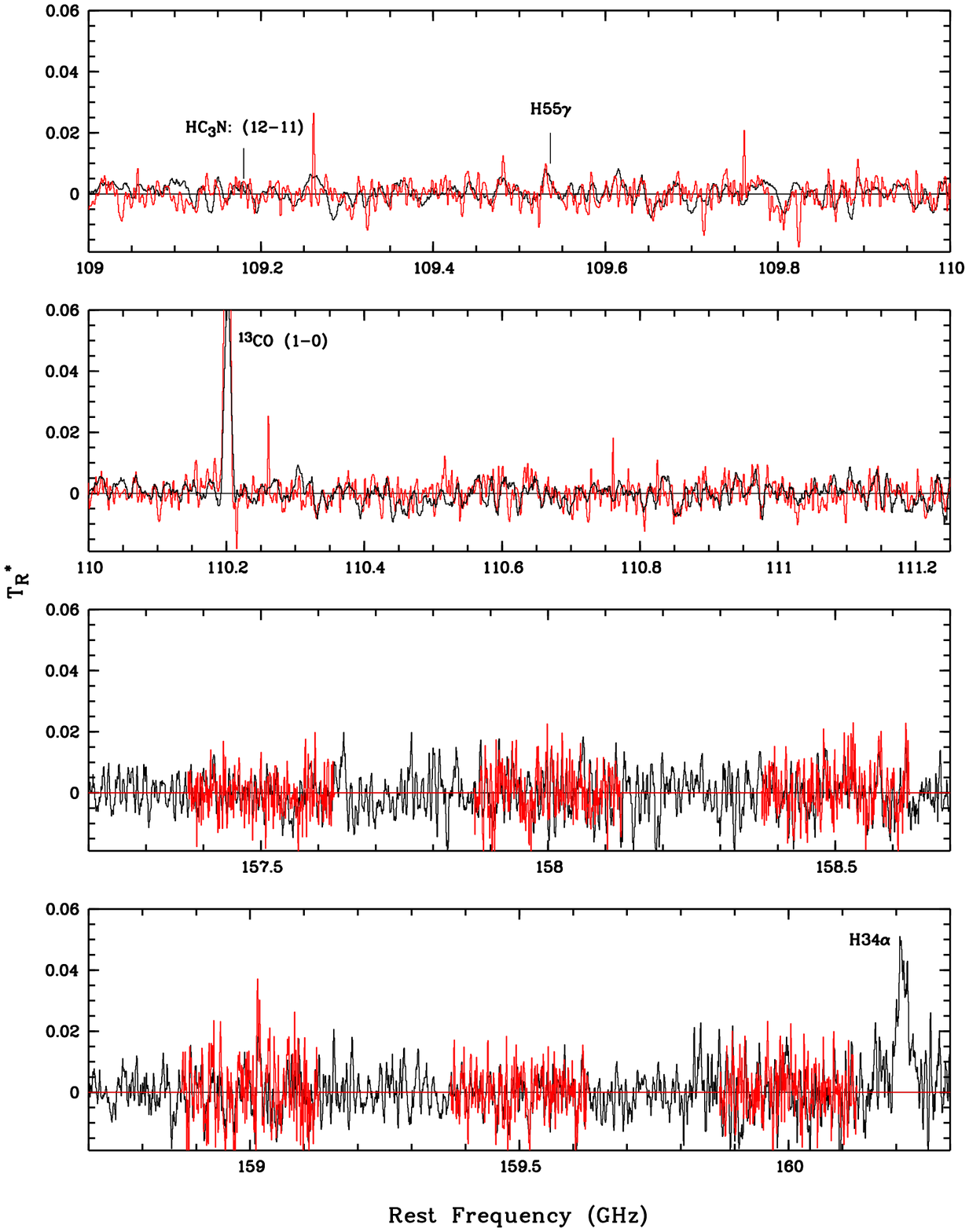}}\\
\centerline{Fig. 3. --- Continued.}

\clearpage

\clearpage
{\plotone{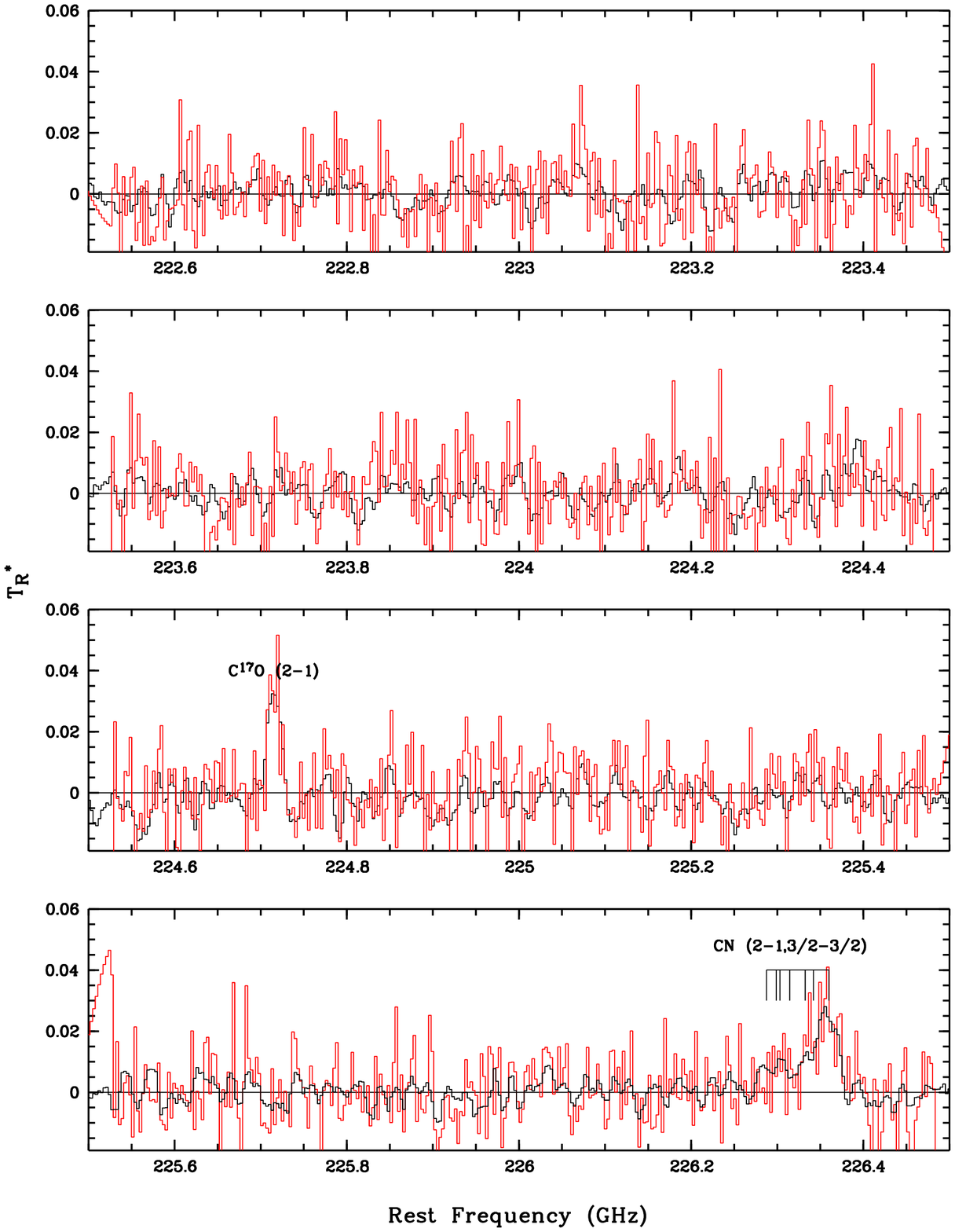}}\\
\centerline{Fig. 4. --- Continued.}
\clearpage
{\plotone{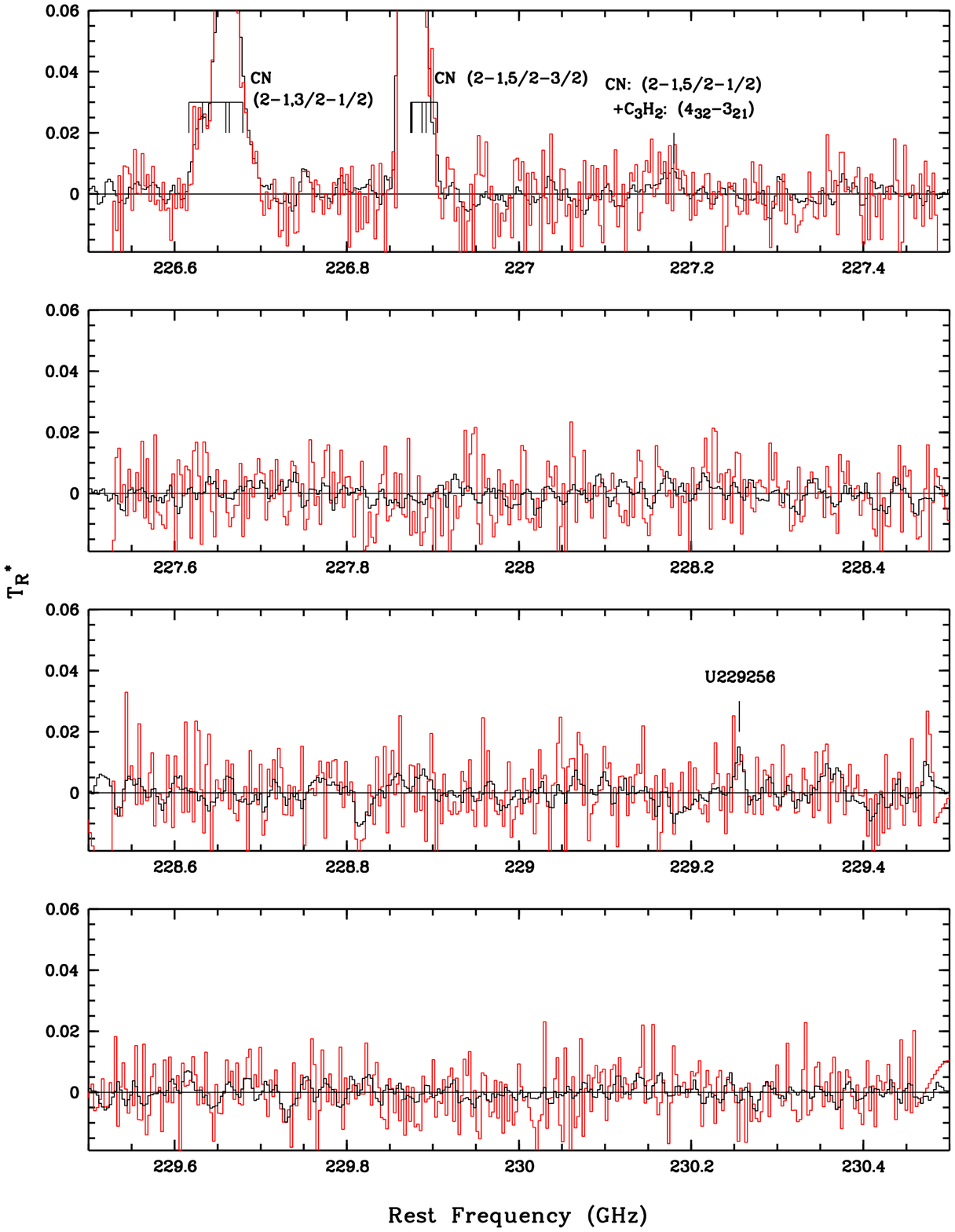}}\\
\centerline{Fig. 4. --- Continued.}
\clearpage
{\plotone{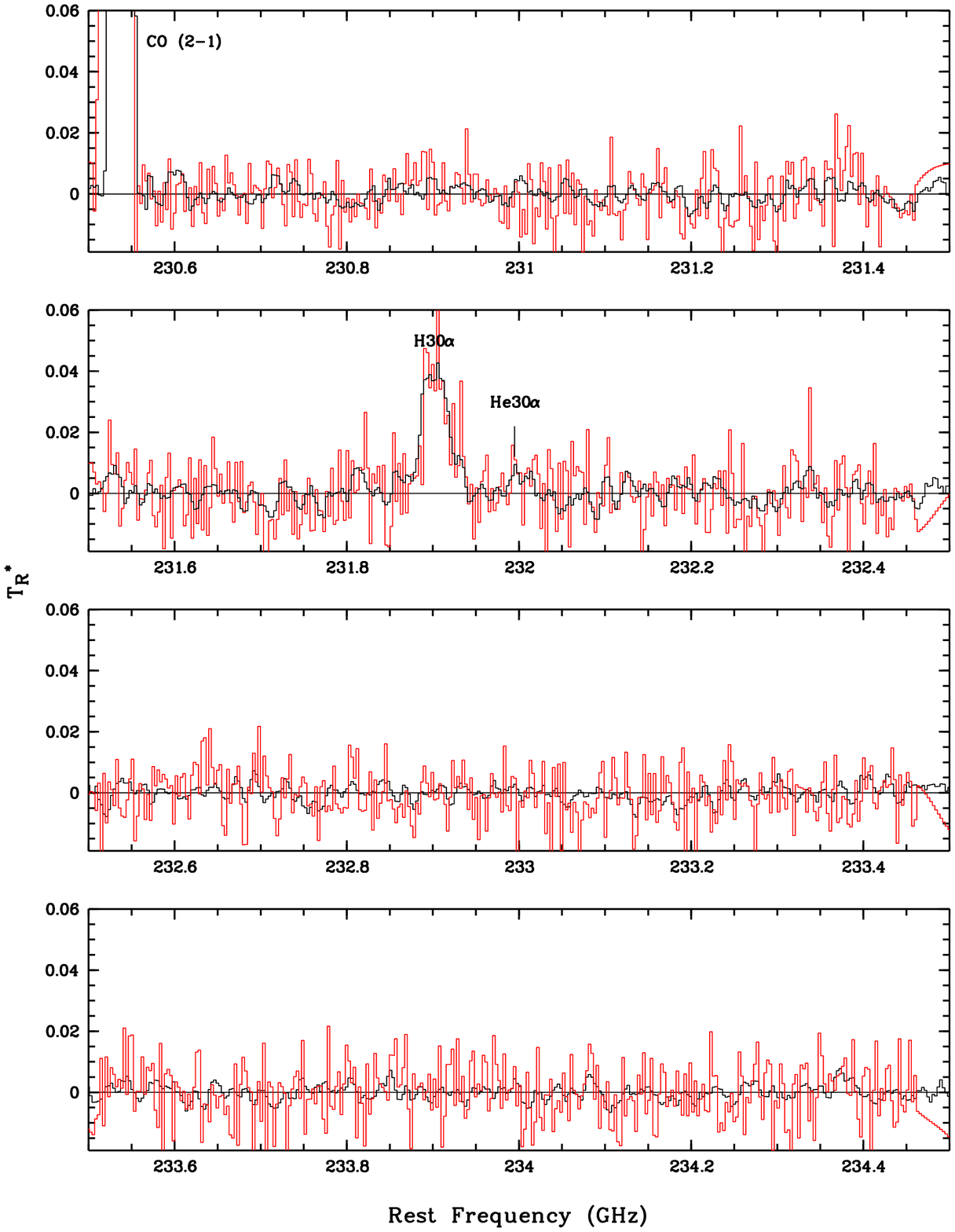}}\\
\centerline{Fig. 4. --- Continued.}
\clearpage
{\plotone{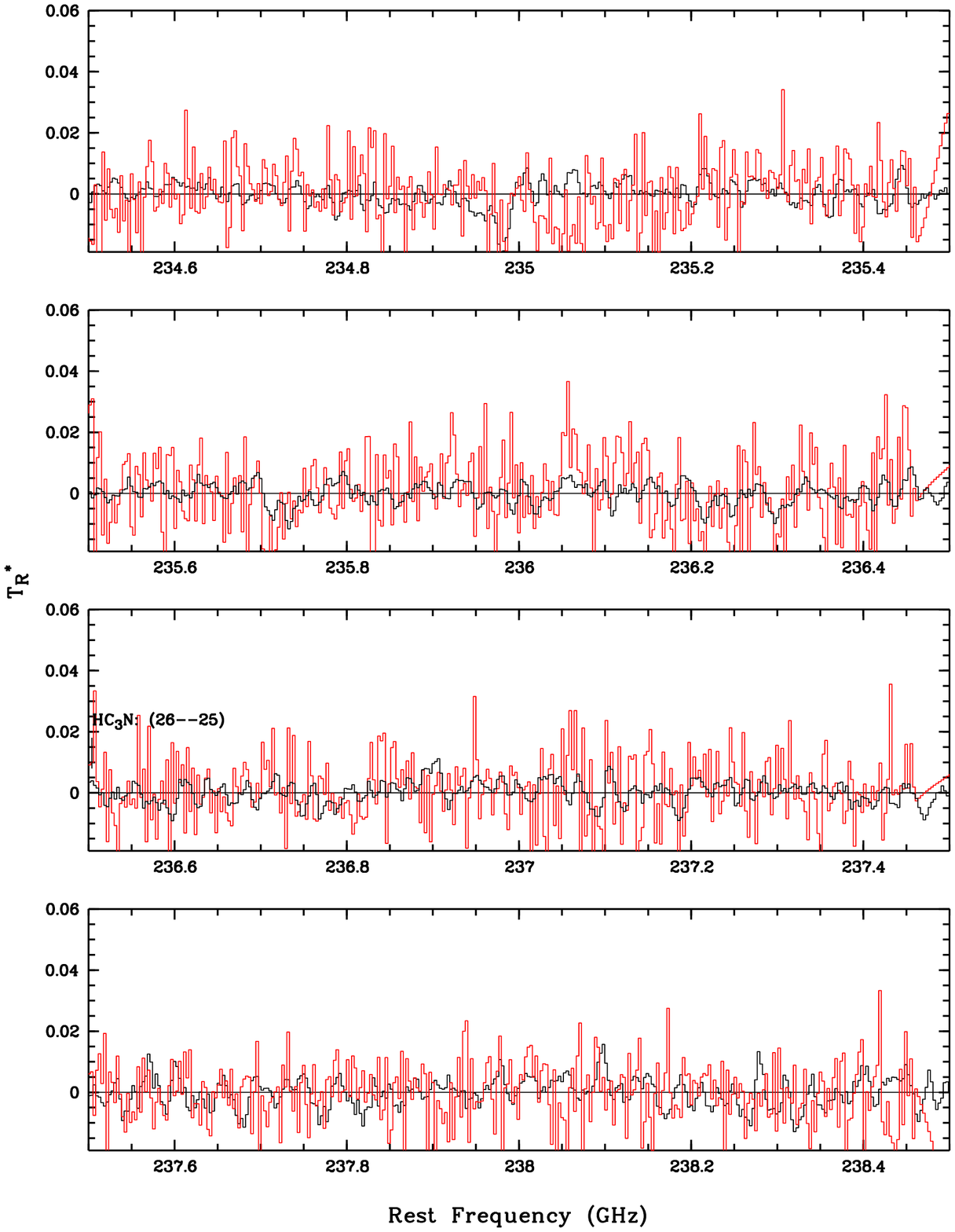}}\\
\centerline{Fig. 4. --- Continued.}
\clearpage
{\plotone{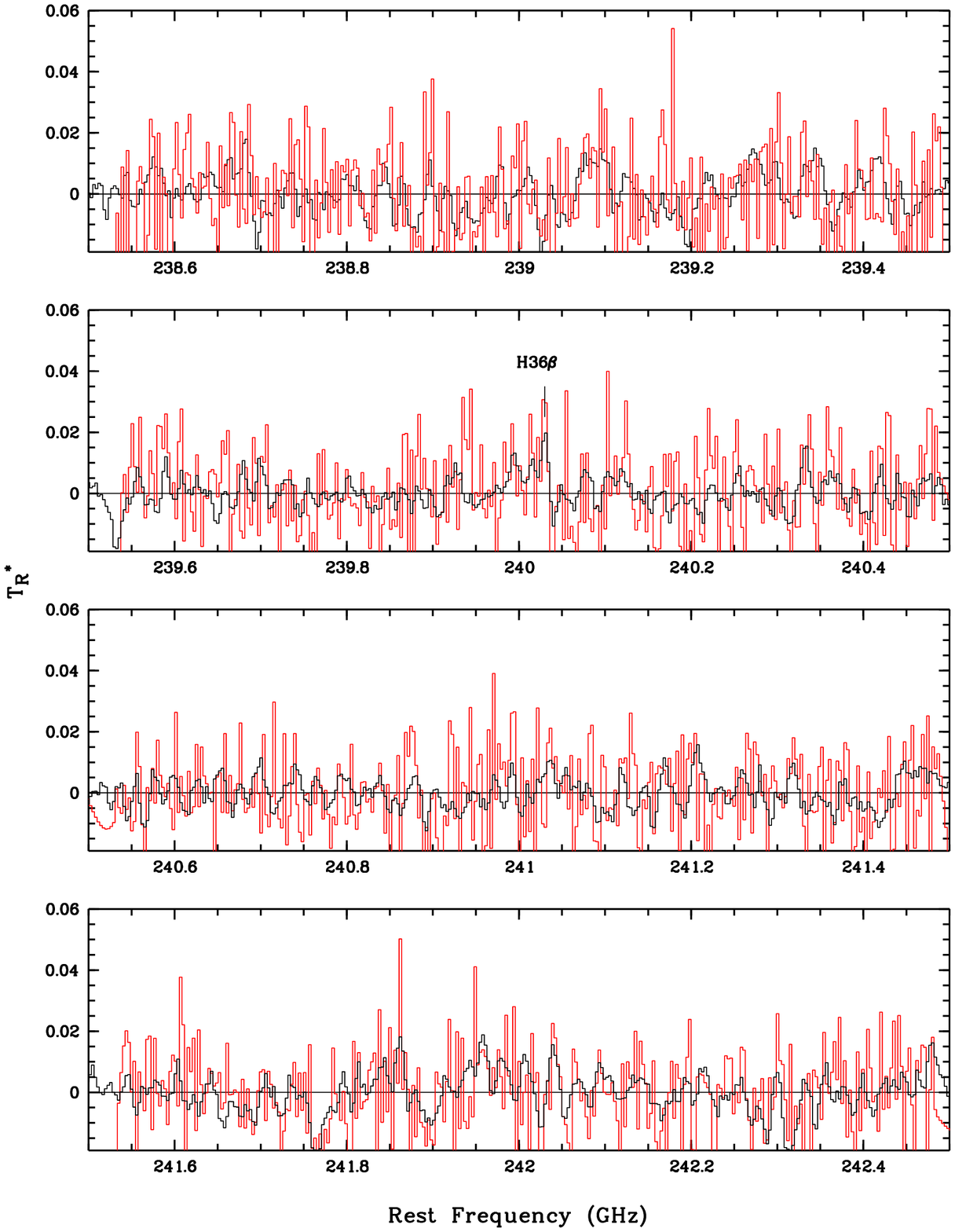}}\\
\centerline{Fig. 4. --- Continued.}
\clearpage
{\plotone{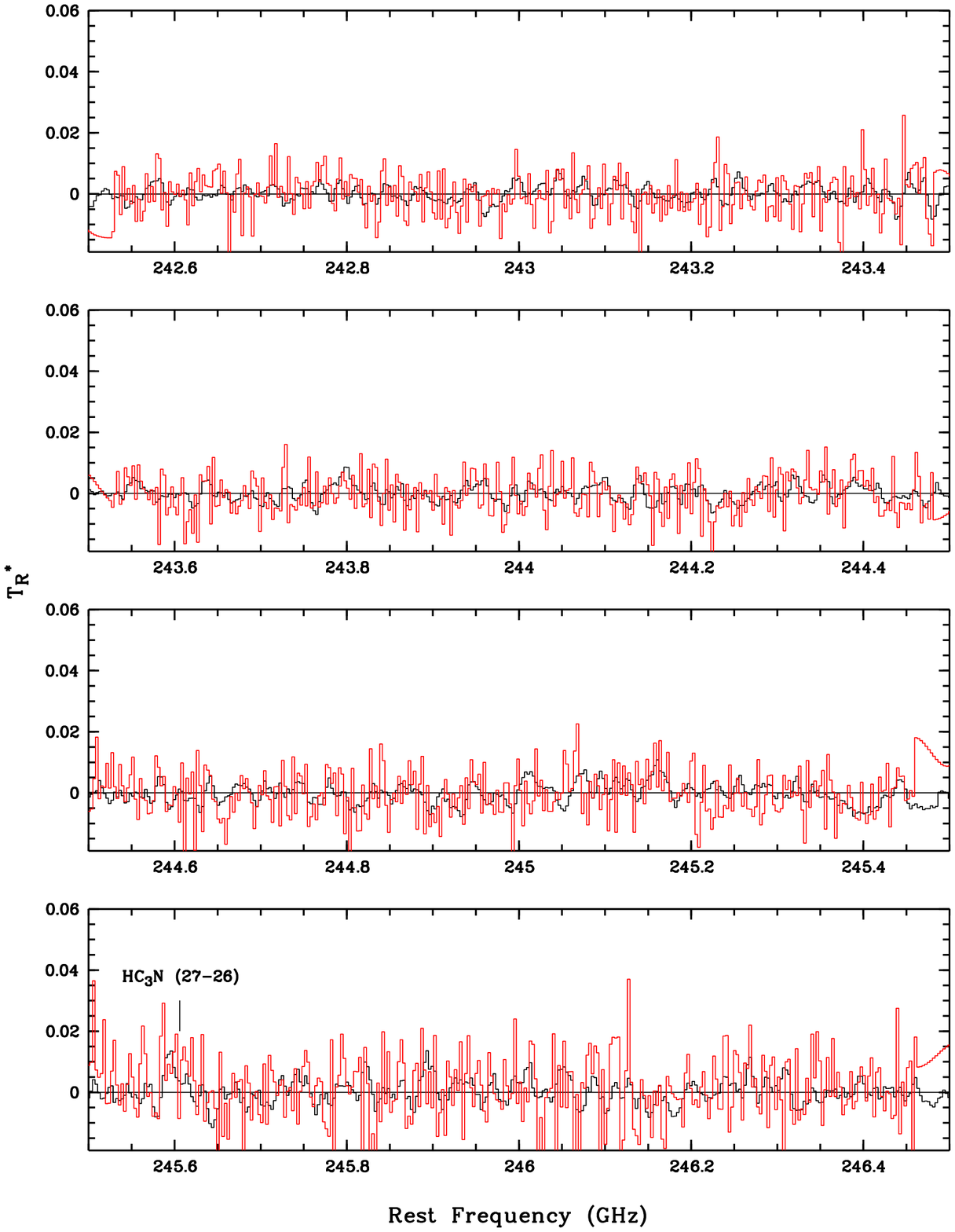}}\\
\centerline{Fig. 4. --- Continued.}
\clearpage
{\plotone{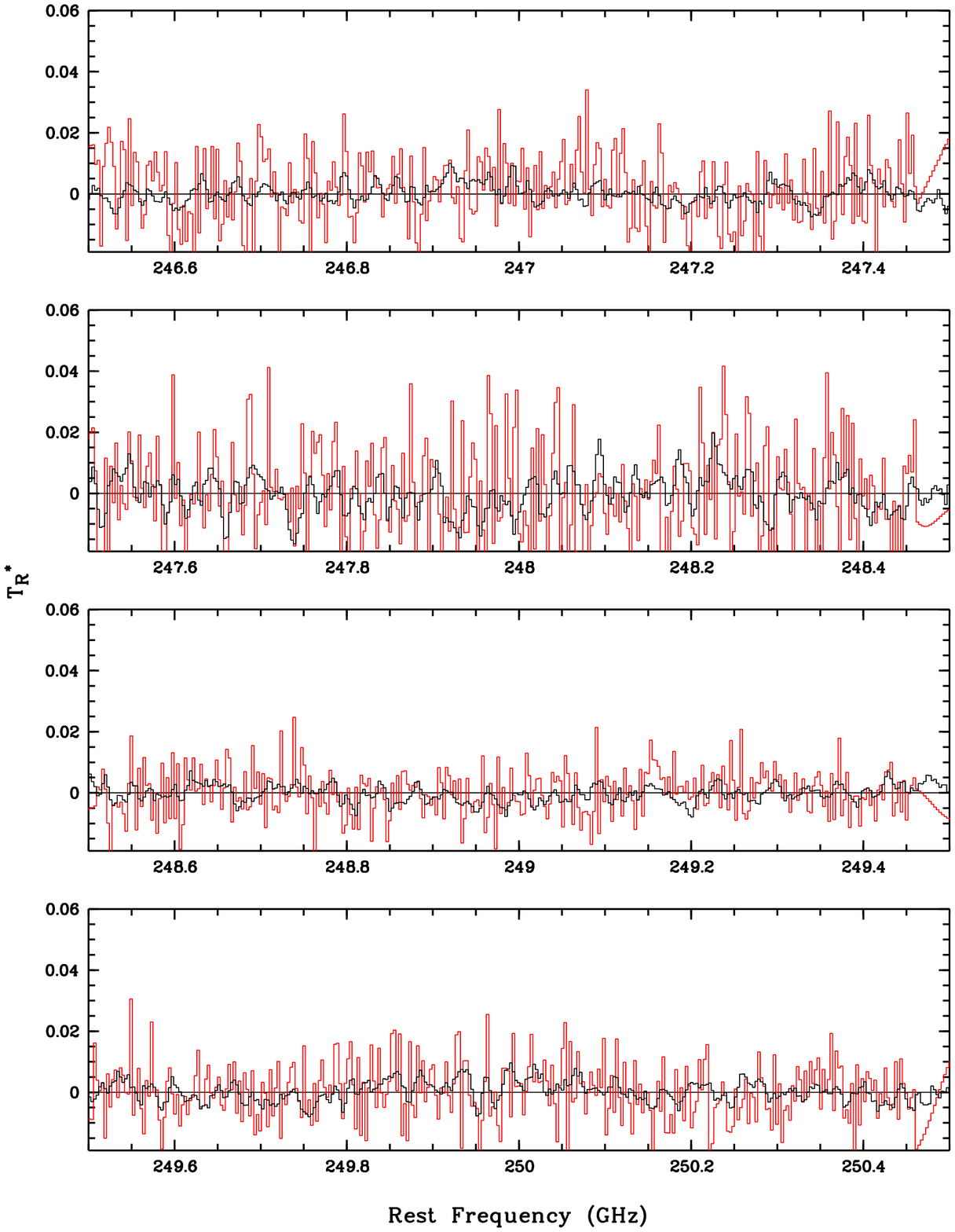}}\\
\centerline{Fig. 4. --- Continued.}
\clearpage
{\plotone{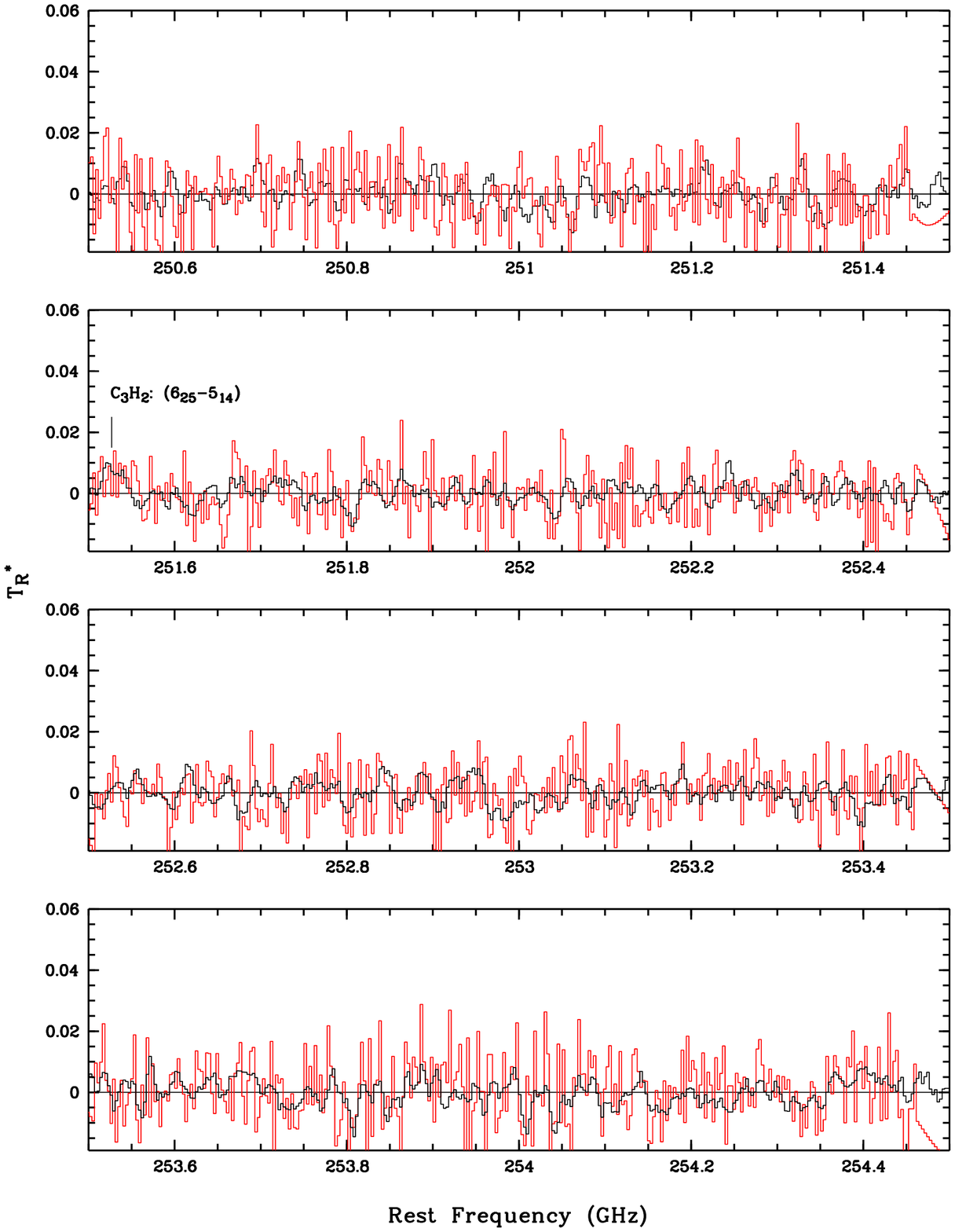}}\\
\centerline{Fig. 4. --- Continued.}
\clearpage
{\plotone{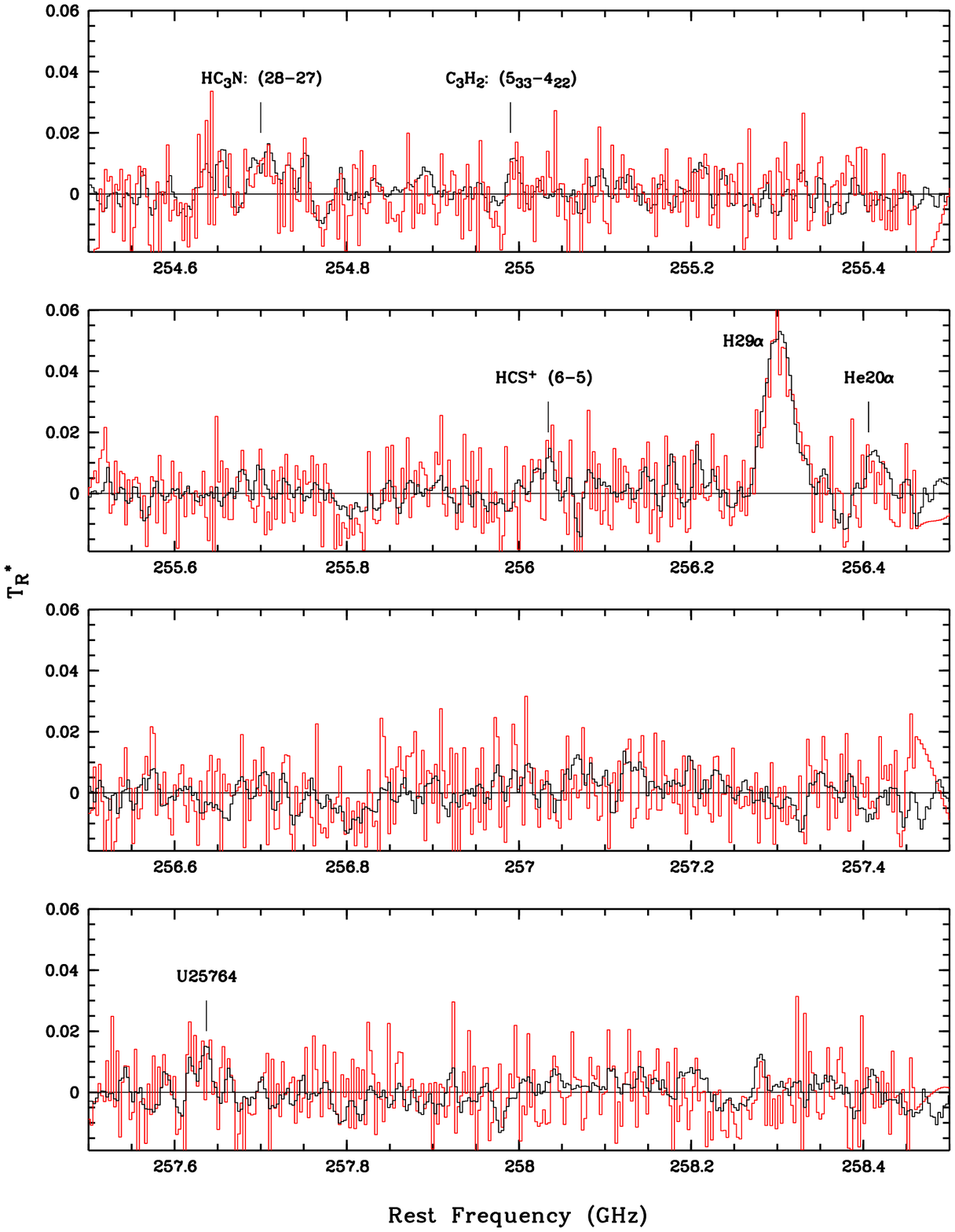}}\\
\centerline{Fig. 4. --- Continued.}
\clearpage
{\plotone{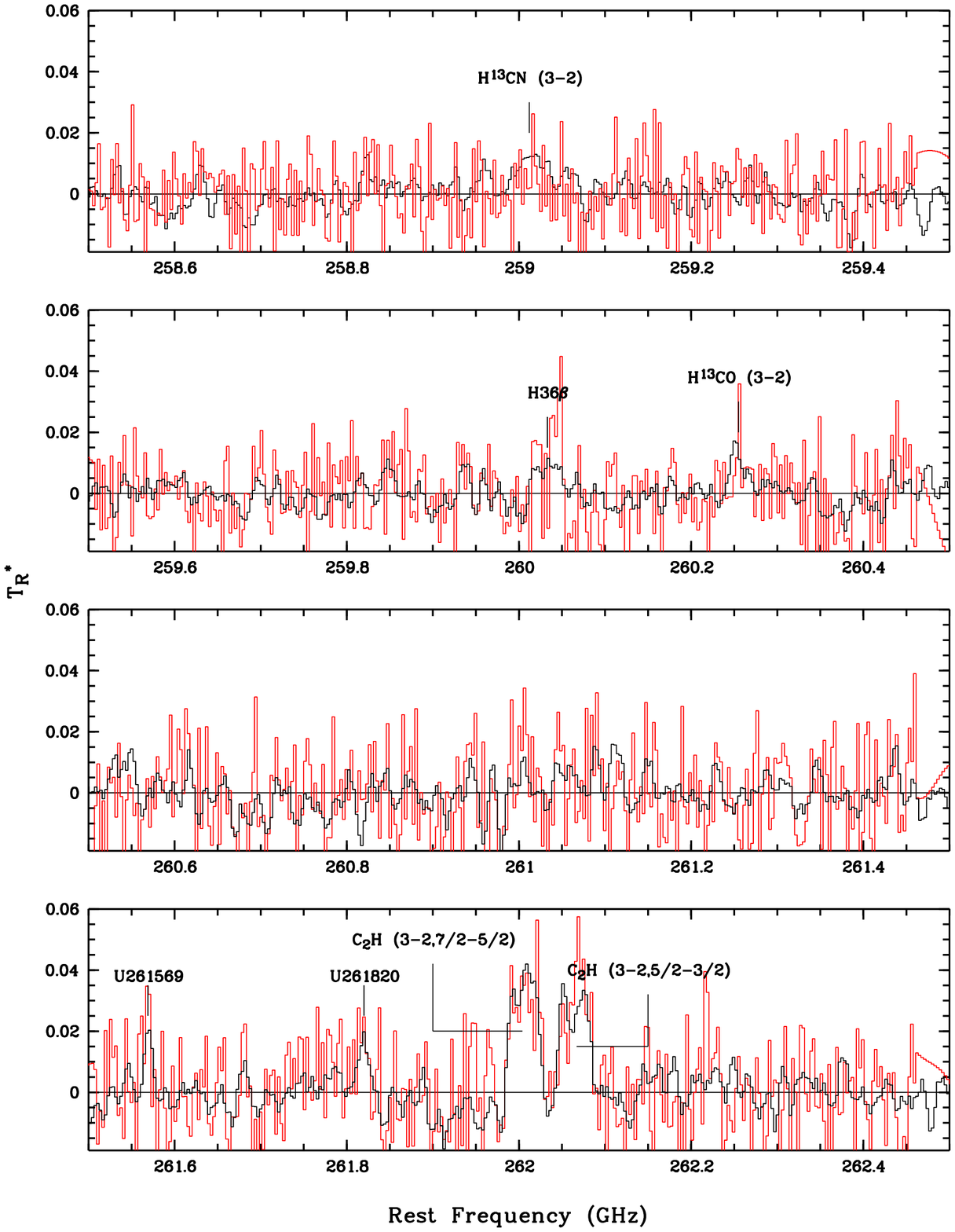}}\\
\centerline{Fig. 4. --- Continued.}
\clearpage
{\plotone{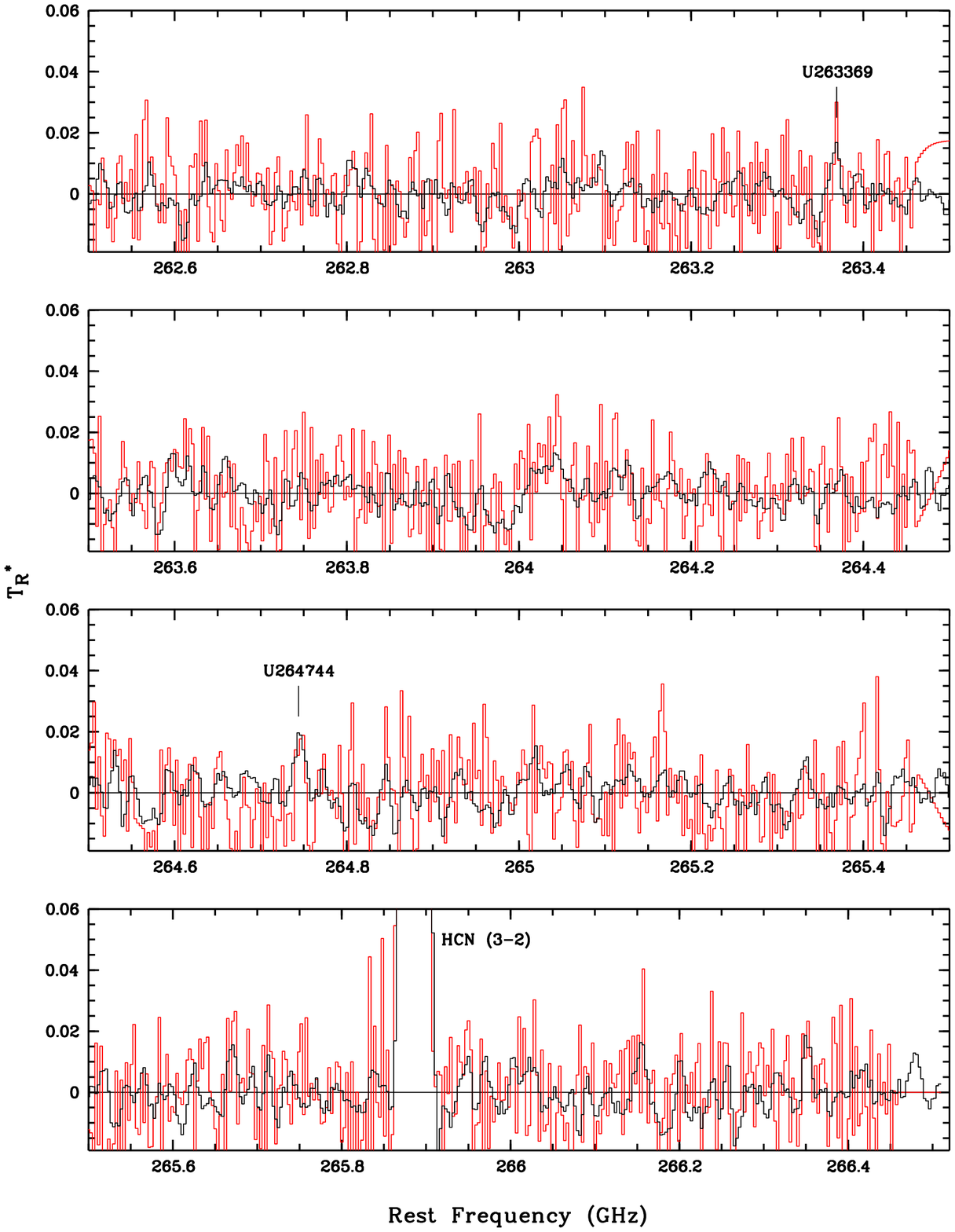}}\\
\centerline{Fig. 4. --- Continued.}

\clearpage
\centerline{\includegraphics[height=20cm]{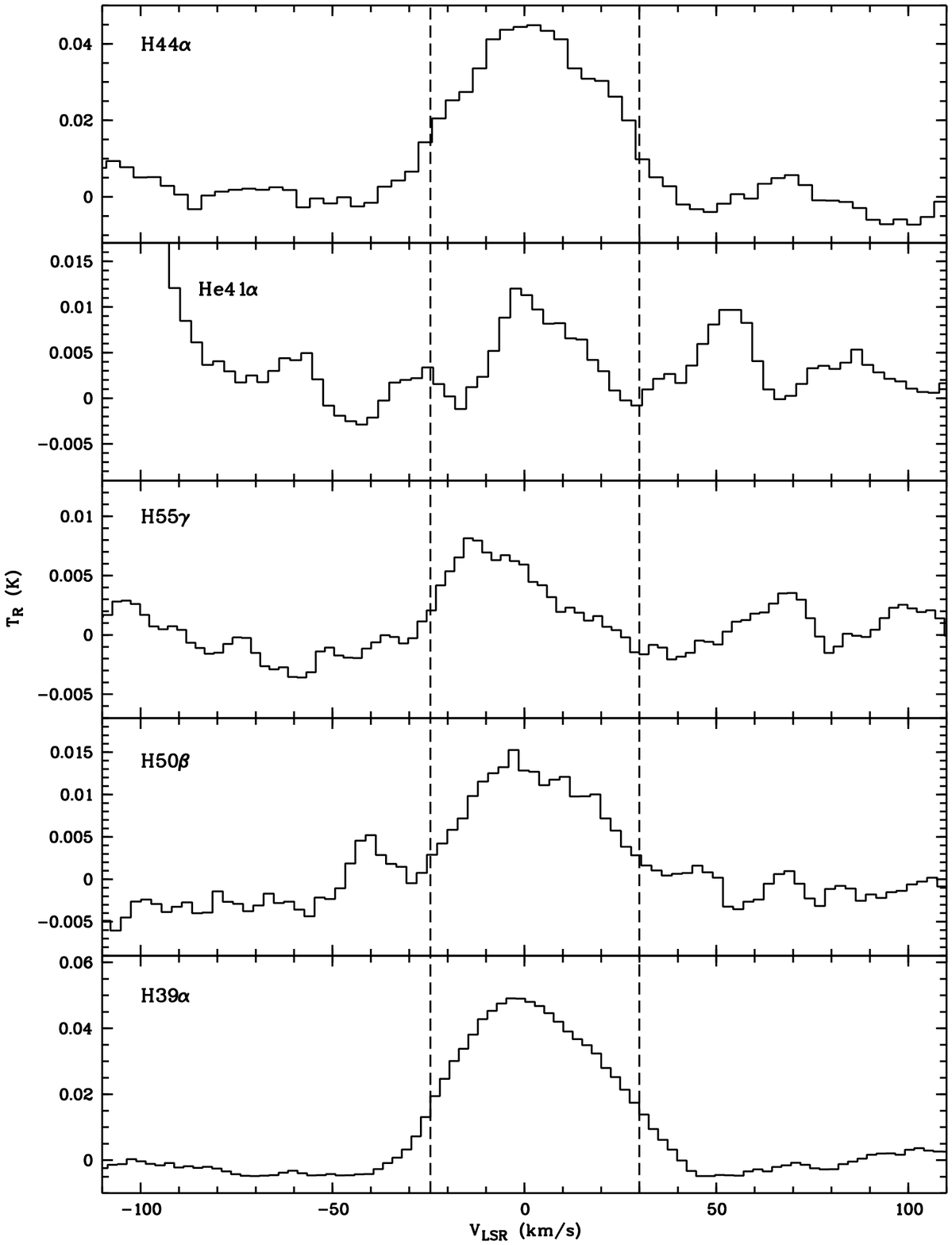}}
\centerline{Fig. 5. --- Continued.}

\clearpage
\clearpage
\centerline{\includegraphics[height=20cm]{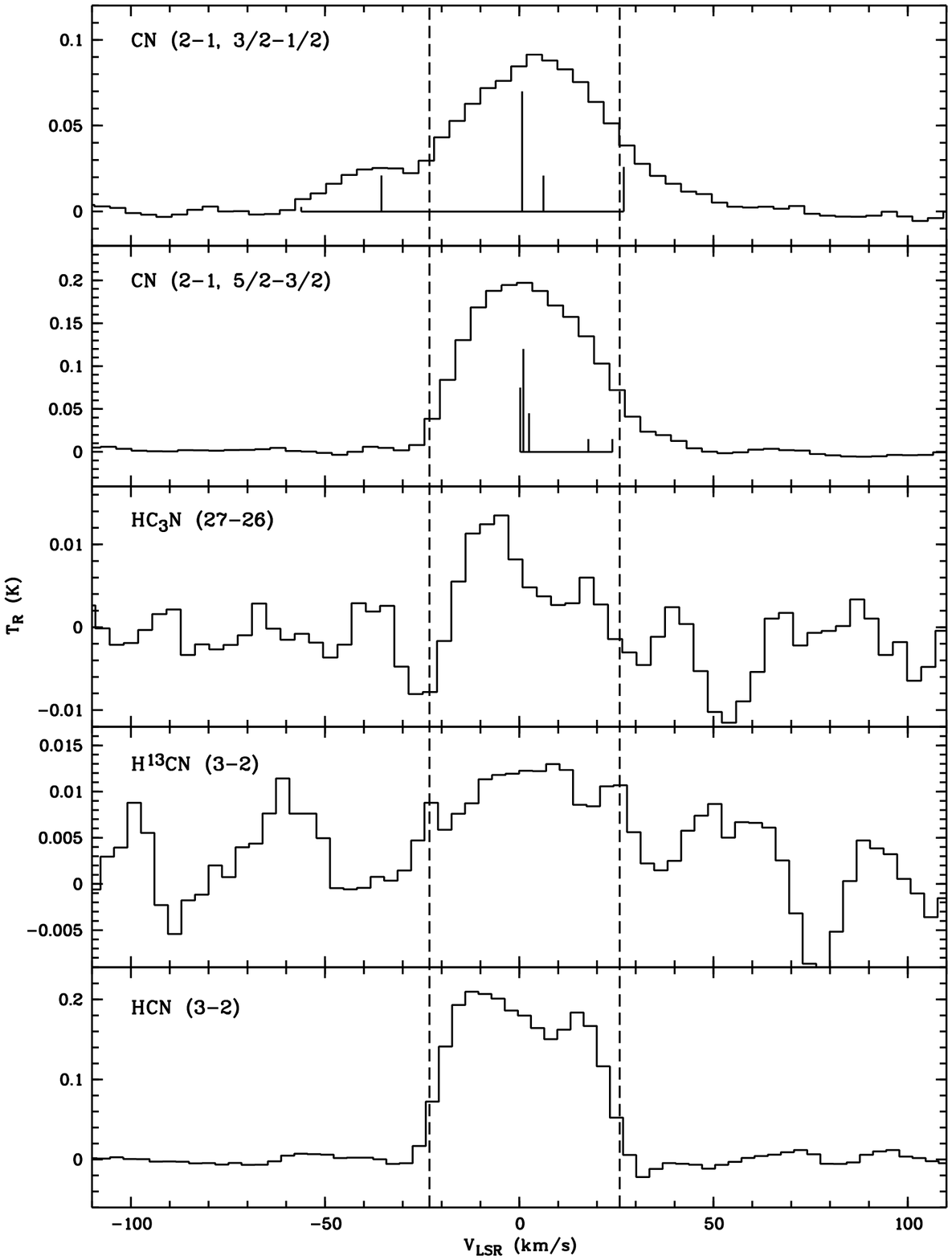}}
\centerline{Fig. 6. --- Continued.}
\clearpage
\centerline{\includegraphics[height=20cm]{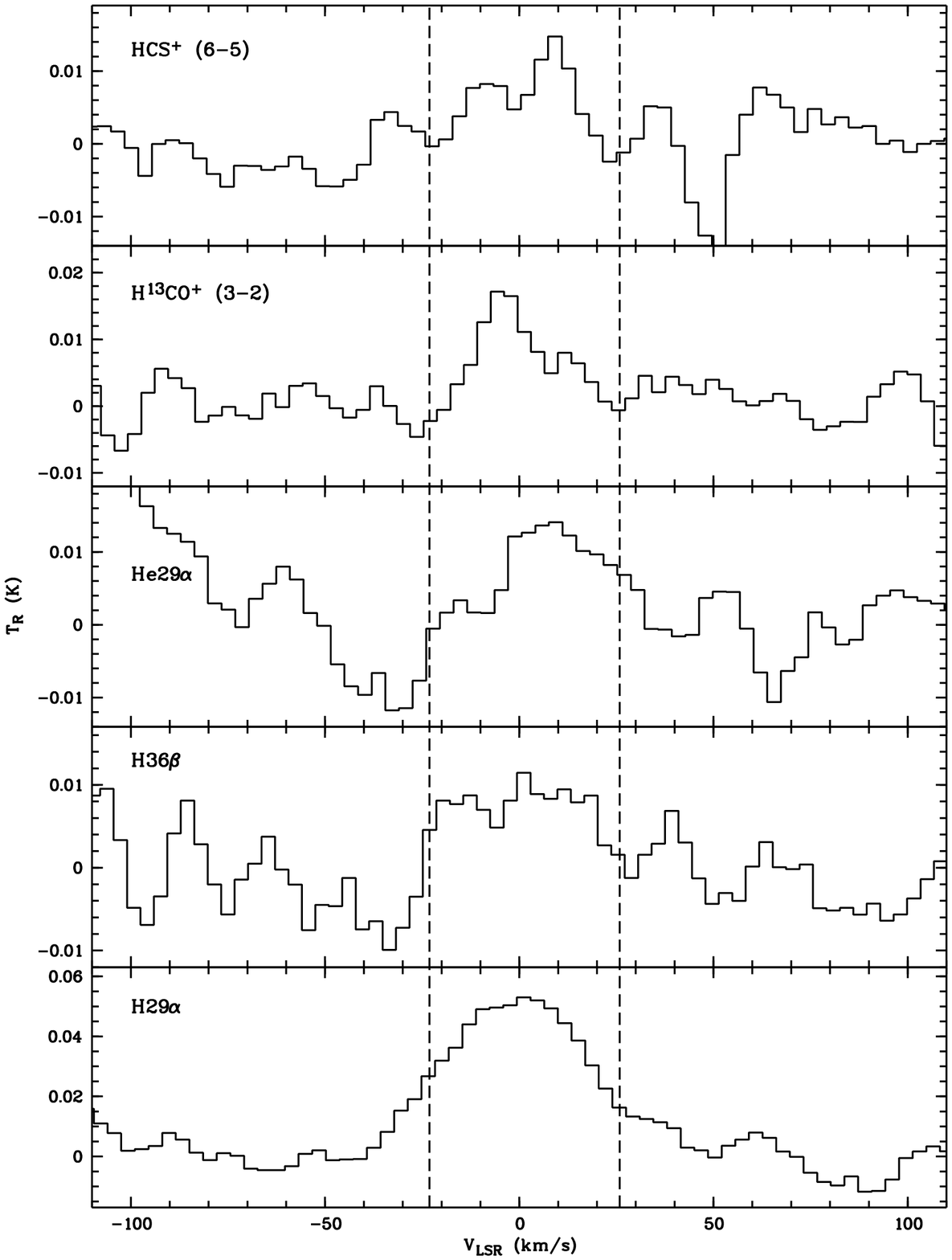}}
\centerline{Fig. 6. --- Continued.}
\clearpage

\end{document}